\documentclass[12pt]{article}
\pdfoutput=1
\usepackage{amssymb,amsmath,amsthm,trig,amsfonts,graphicx}
\usepackage{relsize,ccaption,url,latexsym,epsfig,a4wide}
\numberwithin{equation}{section}

\captionnamefont{\bfseries}

\title{Boundary Giant Magnons and Giant Gravitons}

\author{\\ \Large{A. Ciavarella\footnote{a.m.ciavarella@durham.ac.uk}\;\; and P. Bowcock\footnote{peter.bowcock@durham.ac.uk}}\\ \\ \\
 \normalsize{Centre for Particle Theory}\\ \normalsize{Department of Mathematical Sciences} \\ \normalsize{University of Durham} \\ \normalsize{Durham, DH1 3LE, U.K.  }}

\newcommand{\be}{\begin{equation}}
\newcommand{\ee}{\end{equation}}
\newcommand{\bea}{\begin{eqnarray}}
\newcommand{\eea}{\end{eqnarray}}
\newcommand{\Ra}{\Rightarrow}
\newcommand{\nn}{\nonumber}
\newcommand{\phit}{\dot{\phi}}
\newcommand{\slambda}{\frac{\sqrt{\lambda}}{2\pi}}
\newcommand{\F}{\tilde{F}}

\begin{document}

\maketitle

\abstract{We construct the full set of boundary giant magnons on $\mathbb{R}\times S^{2}$ attached to the maximal $Z=0$ giant graviton by mapping from the general solution to static sine-Gordon theory on the interval and compute the values of $\Delta-J$ at finite $J$, including the leading order corrections when $J$ is large.  We then consider the Born-Infeld theory of the giant graviton itself to construct BIon spike solutions that correspond to the world volume description of the boundary giant magnons at finite $J$.}

\section{Introduction}

It has been known since the work of Pohlmeyer\cite{Pohlmeyer} in the 1970's that there exists a one to one map between the solutions of the O(3) sigma model and those of sine-Gordon theory, the prototypically integrable 1+1 dimensional, Lorentz invariant scalar field theory.  The O(3) sigma model, with the inclusion of the Virasoro constraints, represents a conformal gauge string moving on $\mathbb{R}\times S^{2}$.  The `Pohlmeyer reduction' of more complex sigma models in terms of scalar fields representing quantities built out of invariants under the sigma model symmetry group have recently been constructed\cite{Grigoriev:2007bu} as part of a project to reformulate the theory of the type IIB superstring moving on $AdS_{5}\times S^{5}$ in terms of purely physical degrees of freedom.

The map between integrable scalar field theories and superstrings has proved very fruitful within the $AdS/$CFT correspondence\cite{Maldacena:1997re,Witten:1998qj,Aharony:1999ti} since Hofman and Maldacena's identification\cite{HofmanI} of certain operators of high symmetry and large charge on the conformal field theory side with solitonic solutions of a string moving on an $\mathbb{R}\times S^{2}$ subspace of $AdS_{5}\times S^{5}$ and hence with the solitons of sine-Gordon theory.  Explicit maps between strings moving on higher spheres within $S^{5}$, and in the $AdS_{5}$ part of the space, and solitons of higher scalar field theories\cite{Pohlmeyer:1979ch,D'Auria:1980cx,D'Auria:1979tb,Bakas:1995bm} have since been constructed\cite{Chen:2006gea,Hollowood:2009tw}.

More particularly, Hofman and Maldacena's original identification\cite{HofmanI} took the large $N$ 't Hooft limit  ($N\to\infty$, $g_{{\rm s}}\to 0$ with $\lambda=g_{{\rm s}}N$ fixed) and was between CFT operators composed of an infinite number $J$ of one of the scalar fields $Z$ and a single one of the other scalars $X$ of `momentum' $p$ with a rigid string configuration possessing infinite energy $E$ and angular momentum $J$ about an axis in $S^{5}$. The $AdS$/CFT dictionary relates the conformal dimensional $\Delta$ of a CFT operator to the string energy $E$ while the R-charges of the CFT correspond to charges carried by the string due to the isometries of the compact space on which they move, hence relating the angular momentum of the string to R-charge.  The momentum $p$ is kept fixed as $E$ and $J\to \infty$ but the resulting states have the following finite differences that can be obtained from the constraints imposed by the superconformal algebra alone\cite{Beisert:2005tm} and that match at large $\lambda$ where the classical string approach is valid:
\be
(\Delta-J)_{{\rm CFT}}=\sqrt{1+\frac{\lambda}{\pi^{2}}{\rm sin}^{2}\left(\frac{p}{2}\right)},\qquad (E-J)_{{\rm string}}=\frac{\sqrt{\lambda}}{\pi}\bigg|{\rm sin}\left(\frac{p}{2}\right)\bigg|.\nn
\ee

The CFT states can be seen as a single excitation above a BMN state of \cite{Berenstein:2002jq} that itself corresponds to a point-like string moving about the equator of $S^{5}$ at the speed of light, while the string excitation above that consists a rigid semi-circle on an $S^{2}\subset S^{5}$ with its two light-like ends at the equator possessing the properties of two BMN strings; that is two BMN vacua are interpolated between by the extended string representing the single `impurity' scalar $X$ in the CFT operator.  An equivalence between the CFT operators and those of a spin-chain\cite{Minahan:2002ve} led to the CFT operator being referred to as a magnon state while the string configuration was called the giant magnon.  The giant magnon possesses a non-zero net world sheet momentum so that it is technically not a physical string state, reflecting the fact that the CFT operator is left untraced over and is hence not gauge invariant.  However, by adding giant magnons together so as to obtain net zero world sheet momentum we can create a physical closed string.
\newline

The description of the bosonic string on $\mathbb{R}\times S^{2}$ in the Pohlmeyer reduced theory is simply sine-Gordon theory.  Under this map the string world sheet and sine-Gordon theory share the very same spacial variable $x$ and time variable $t$.\footnote{We take the static gauge such that the world sheet time variable $\tau$ is identified with the global time variable $t$ of $AdS_{5}$.}  The Lagrangian density of sine-Gordon theory for a single scalar field $\varphi\in\mathbb{R}$ is
\be
\mathcal{L}_{{\rm SG}}=\frac{1}{2}\dot{\varphi}^{2}-\frac{1}{2}\varphi'^{2}+g{\rm cos}\left(\beta\varphi\right)
\ee
where a dot and a prime denote time and space derivatives respectively.  After rescaling of the field and coupling constant the equation of motion derived from this Lagrangian density is the sine-Gordon equation
\be
\ddot{\varphi}-\varphi''=-{\rm sin}(\varphi).
\ee
In the bulk ($-\infty<x<+\infty$) this equation has the well known $n$-soliton solutions in the form of kinks and anti-kinks (and any mixture thereof) that smoothly interpolate between the vacuum values $\varphi=2m\pi$, $m\in\mathbb{Z}$.  At early and late times the solution is approximately the superposition of $n$ independently boosted solitons, with the condition that no two solitons share the same velocity.  A configuration of $n$ scattering kinks and anti-kinks here becomes $n$ scattering giant magnonic excitations of the string world sheet on $\mathbb{R}\times S^{2}$ with giant magnons corresponding to kinks lying on one hemisphere and those corresponding to anti-kinks lying on the other.

The sine-Gordon field is defined through the string fields by
\be
{\rm cos}(\varphi)=\dot{\vec{X}}^{2}-\vec{X}'^{2}\label{sGdef1}
\ee
where for e.g. $\dot{\vec{X}}^{2}\equiv\dot{\vec{X}}\cdot\dot{\vec{X}}\equiv \dot{X}_{i}\dot{X}^{i}$, $i$ running over spacial target-space values.  Due to the derivatives in this definition the string fields are nonlocal functions of the sine-Gordon field so that it is in general a very difficult task to translate a given solution of sine-Gordon theory to its corresponding string solution.  Certain simple cases are amenable however.

In the conformal gauge a string has a constant energy density proportional to $\dot{\vec{X}}^{2}+\vec{X}'^{2}=1$, ensured by one of the Virasoro constraints resulting from the choice of gauge.  The above definition of $\varphi$ is equivalently
\be
\dot{\vec{X}}^{2}={\rm cos}^{2}\left(\frac{\varphi}{2}\right),\qquad \vec{X}'^{2}={\rm sin}^{2}\left(\frac{\varphi}{2}\right)\label{sGdef2}
\ee
which clearly automatically satisfies this physicality condition. Due to the constant energy density, our giant magnons, which are of infinite energy, possess a spacial variable with infinite range: this decompactifying limit is part of what allows us to consider world sheets with a net momentum and draws the correspondence between giant magnons and the \emph{bulk} sine-Gordon theory (the string end points being pushed off to $x=\pm\infty$).  That the theories share the time variable leads to the possibility of calculating magnon scattering phase shifts from soliton-soliton scattering time delays.

The map needn't be restricted to the bulk theory however.  In \cite{HofmanII} Hofman and Maldacena studied giant magnons scattering on the world sheet of strings with one end point attached to a giant graviton\cite{McGreevy}, a D3 brane wrapped on an $S^{3}\subset S^{5}$.  They considered three simple and physically distinct orientations of the giant graviton in $S^{5}$ with respect to the $S^{2}$ on which the string moves and, taking again the large $J$ limit so as to decouple the two endpoints of the spin-chain or string, these mapped to three simple boundary conditions in \emph{boundary} sine-Gordon theory\cite{Saleur:1994yh}.  The giant gravitons considered were the so called maximal variety, meaning that the radius of the $S^{3}$ wrapped by the gravitons is taken to be its maximal value, equal to that of the $S^{5}$.  Phase shifts could again be calculated by reference to the time delays experienced by a soliton reflecting off a boundary in the sine-Gordon picture and successfully compared with those obtained in the CFT.
\newline

Taking $J$ finite leads to finite volume effects on the world sheet and in the spin-chain pictures alike. Giant magnons at finite $J$ were first addressed by Arutyunov, Frolov and Zamaklar\cite{Arutyunov:2006gs} where the string solutions were found and used to calculate corrections to the finite quantity $\Delta-J$.  In particular it was found that the leading corrections are exponentially small in $J$,
\be
\Delta-J\approx\frac{\sqrt{\lambda}}{\pi}{\rm sin}\left(\frac{p}{2}\right)\left\{1-\frac{4}{e^{2}}{\rm sin}^{2}\left(\frac{p}{2}\right)e^{-\frac{2\pi}{\sqrt{\lambda}}{\rm cosec}\left(\frac{p}{2}\right)J}\right\}.\label{finiteJmagnon}
\ee

For a given $p$ the individual giant magnons do not obey physical end point condition and are not rigid configurations.  However, we may again attach a number of such strings end to end in order to create a physical state.  Exceptionally, at $p=\pi$ when the magnon passes over the pole of $S^{2}$, the string is rigid and perfectly physical and is simply half of the spinning string\cite{Gubser:2002tv} of angular momentum $J$ on $\mathbb{R}\times S^{2}$.

A very similar string solution to this is the boundary giant magnon attached to a ``$Z=0$''\footnote{The $Z$ is one of the three complex coordinates used to embed $S^{5}$ in $\mathbb{R}^{6}$.  Setting one of these complex coordinate equal to zero then decides the orientation of the brane in $S^{3}$.} giant graviton studied by Bak\cite{Bak} in an examination of string zero modes.  It has \emph{both} end points on the giant graviton, has no giant magnon excitations travelling on the world sheet, and is folded back on itself with a cusp point that as $J\to\infty$ stretches down to the equator to form two halves of a single giant magnon with $p=\pi$.  At $J$ large but finite it was found that
\be
\Delta-J\approx\frac{\sqrt{\lambda}}{\pi}\left\{1-\frac{4}{e^{2}}e^{-\frac{2\pi}{\sqrt{\lambda}}J}\right\}\label{finiteJBak}
\ee
which is precisely the result \eqref{finiteJmagnon} with $p=\pi$.  This boundary giant magnon is precisely two coincident halves of a normal giant magnon. 
\newline

From a different perspective the giant magnon has been studied as an object appearing on the world volume of a giant graviton\cite{Sadri:2003mx,Hirano} as a solitonic solution of the Born-Infeld theory (plus Chern-Simons potential) analogous to the original BIon spike solutions\cite{Callan:1997kz,Gibbons:1997xz} describing fundamental strings attaching to D-branes in flat space. The presence of electrical charges on the world volume corresponds to the end points of F-strings ending on the D-brane and the solitonic string solution requires both scalar and electric profiles that conspire perfectly so as to create a BPS object.  The giant graviton is a compact object and so cannot admit of solutions with a net electric charge, though dipole (and higher pole solutions) are possible.  The giant magnon satisfyingly appears as two equal and oppositely orientated BIon spikes emerging from the poles of a single giant graviton.  In \cite{Hirano} the dipole solution was named the ``fat magnon" while it's multipole counterpart studied on a plane-wave background in \cite{Sadri:2003mx} was named the ``giant hedge-hog".
\newline

In this paper we take another look at boundary giant magnons at finite $J$ attached to ``$Z=0$'' giant gravitons.  Our purpose is to find the boundary giant magnons on $\mathbb{R}\times S^{2}$ in the Pohlmeyer reduced picture, i.e. in sine-Gordon theory, as well as in the world volume theory as BIon spike solutions.  Finite $J$ means looking at sine-Gordon theory on the interval, and we will be able to set the time derivative of $\varphi$ to zero.

Recalling that the end points of the giant magnons move at the speed of light, we require that the angular variable $\phi$, which being a transverse direction to the giant graviton appears as a scalar field on its world volume, satisfies $\phit=1$.  Taking finite $J$ we move away from $\phit=1$.  The finite $J$ corrections to $\Delta-J$ calculated for the giant magnon itself\cite{Arutyunov:2006gs}, as well as the boundary magnon in \cite{Bak}, each took $\phit\geq 1$.  In this paper we will examine both $\phit\geq 1$ and $0\leq\phit\leq 1$ for which the solutions display qualitatively different behaviour.
\newline

The layout of the paper is as follows. In section 2 we present the general solutions to static sine-Gordon theory on the interval and take focus on the boundary conditions relevant to maximal giant gravitons.  In section 3 we map these solutions to the string solutions and calculate the quantity $\Delta-J$ for any $\phit$.  In section 4 we study the world volume description of the boundary giant magnons at finite $J$ to rediscover the strings of the previous section.  Section 5 is the conclusion.

\section{Static Sine-Gordon Solutions on the Interval}

In order to produce the boundary giant magnon string solutions at finite $J$ we in this section produce their Pohlmeyer reduced versions from the general solution to static sine-Gordon theory on the interval.  Hence we take all time derivatives of $\varphi$ to be zero, and to place the theory on the interval we will demand that $0\leq x\leq L$.  As we have set the time derivative of $\varphi$ to zero we will have two Dirichlet boundary conditions to consider at $x=0$ and $L$.  We will focus on those for which $\varphi(0)=\varphi(L)+2m\pi$, $m\in\mathbb{Z}$ as in the static case this corresponds to open strings that end on the same D-brane.  The same solutions described here have been previously deployed\cite{Bajnok:2004ka,Mussardo:2004vn} in semiclassical analyses of sine-Gordon theory on the interval.

For static solutions on the interval we take $\dot{\varphi}=\ddot{\varphi}=0$ and will therefore need to find solutions to
\be
\frac{{\rm d}^{2}\varphi}{{\rm d}x^{2}}={\rm sin}(\varphi),\qquad 0\leq x \leq L.\label{staticSGEOM}
\ee
The equation is separated and integrated once as
\bea
\frac{1}{2}{\rm d}\varphi'^{2}&=&{\rm d}\varphi~{\rm sin}(\varphi)\nonumber\\
\Rightarrow \varphi'^{2}&=&2(c-{\rm cos}(\varphi)).
\eea
At this point it is clear that we will have two types of solution depending upon the value of the constant of integration $c$;
\begin{itemize}
\item	if $c>1$ then at no point can $\varphi'=0$
\item	if $|c|\leq 1$ we may obtain $\varphi'=0$
\end{itemize}
while for $c<-1$ there is no real solution.

\subsection{$c> 1$ Solution}
We must solve
\be
\int\frac{{\rm d}\varphi}{\sqrt{c-{\rm cos}(\varphi)}}=\sqrt{2}~{\rm d}x,\qquad c>1.
\ee
The solution is found in terms of an elliptic integral of the first kind to be
\bea
\frac{2}{\sqrt{c+1}}F(\delta,k)=\sqrt{2}(x-x_{0})&\\
\delta={\rm arcsin}\left(\sqrt{\frac{(c+1)(1-{\rm cos}(\varphi))}{2(c-{\rm cos}(\varphi))}}\right),&\qquad k=\sqrt{\frac{2}{1+c}}\nonumber
\eea
or
\be
F({\rm am}(u,k),k)=u,\qquad u=\frac{x-x_{0}}{k}
\ee
${\rm am}(u,k)$ being the Jacobi amplitude function, with $k$ the elliptic modulus satisfying
\be
0\leq k < 1.\nonumber
\ee
This solution is in implicit form however so to make it explicit in $\varphi$ we write
\bea
{\rm am}(u,k)&=&{\rm arcsin}\left(\frac{1}{k}\sqrt{\frac{(1-{\rm cos}(\varphi))}{(c-{\rm cos}(\varphi))}}\right)\nonumber\\
\Rightarrow {\rm sin}({\rm am}(u,k))&\equiv&{\rm sn}(u,k)=\frac{1}{k}\sqrt{\frac{(1-{\rm cos}(\varphi))}{(c-{\rm cos}(\varphi))}}.
\eea
With some rearrangement, including substituting in for $c$ in terms of the modulus $k$, we find
\bea
{\rm cos}(\varphi)&=&2~\frac{{\rm cn}^{2}(u,k)}{{\rm dn}^{2}(u,k)}-1=2{\rm sn}^{2}(u+K,k)-1\nonumber\\
\Rightarrow {\rm cos}^{2}\left(\frac{\varphi}{2}\right)&=&{\rm sn}^{2}(u+K,k).\nonumber
\eea
where $K=K(k)$ is the complete elliptic integral of the first kind.  We get finally
\be
{\rm sin}^{2}\left(\frac{\varphi}{2}\pm\frac{\pi}{2}\right)={\rm sin}^{2}({\rm am}(u+K,k))\nonumber
\ee
or
\be
\varphi=\pm\pi\pm 2~{\rm am}\left(\frac{x-x_{0}}{k}+K,k\right)\label{sGsol1}
\ee
with signs uncorrelated.  It can be checked that this satisfies the static sine-Gordon equation \eqref{staticSGEOM}.

\subsection{$|c|\leq 1$ Solution}
Now we must solve
\be
\int\frac{{\rm d}\varphi}{\sqrt{c-{\rm cos}(\varphi)}}=\sqrt{2}~{\rm d}x,\qquad |c|\leq 1,~c\neq -1.
\ee
This time the solution is of the form
\bea
-\sqrt{2}F\left(\gamma,q\right)=\sqrt{2}(x-x_{0}),&\\
\gamma={\rm arcsin}\left(\sqrt{\frac{(1-{\rm sin}\left(\varphi-\frac{\pi}{2}\right))}{1+c}}\right),&\qquad q=\sqrt{\frac{1+c}{2}}\nn
\eea
or
\be
F\left({\rm am}\left(u',\frac{1}{k}\right),\frac{1}{k}\right)=-(x-x_{0}),\qquad u'=-(x-x_{0}).
\ee
The previously defined constant $k$ now takes the values $1\leq k$ and we proceed to an explicit form of the solution.
\bea
{\rm am}(u',k^{-1})&=&{\rm arcsin}\left(\sqrt{\frac{(1-{\rm sin}\left(\varphi-\frac{\pi}{2}\right))}{1+c}}\right)\nonumber\\
\Rightarrow {\rm sn}(u',k^{-1})&=&\sqrt{\frac{(1-{\rm sin}\left(\varphi-\frac{\pi}{2}\right))}{1+c}}\nonumber\\
&=&\frac{k}{\sqrt{2}}\sqrt{1+{\rm cos}(\varphi)}\nonumber\\
\Rightarrow {\rm sn}\left(u,\frac{1}{k}\right)&=&\pm k~{\rm cos}\left(\frac{\varphi}{2}\right),\qquad u=(x-x_{0})
\eea
where in the last step we absorbed the negative sign in the argument $u'$ into the free choice of signs on the right hand side.  Finally then,
\be
\varphi=2~{\rm arccos}\left(\pm\frac{1}{k}~{\rm sn}\left(x-x_{0},\frac{1}{k}\right)\right).\label{sGsol2}
\ee
Again, it can be checked that this satisfies the equation of motion \eqref{staticSGEOM}.

\subsection{Boundary Conditions}

Our two qualitatively different general solutions to static sine-Gordon theory on the interval are
\bea
c> 1,~(0\leq k< 1),\qquad \varphi&=&\pm\pi\pm2~{\rm am}\left(\frac{x-x_{0}}{k}+K,k\right)\label{phisol1}\\
|c|\leq 1,~~~~~~(1\leq k),\qquad \varphi&=&2~{\rm arccos}\left(\pm\frac{1}{k}~{\rm sn}\left(x-x_{0},\frac{1}{k}\right)\right)\label{phisol2}
\eea
with $0\leq x\leq L$ and $c\neq-1$ in \eqref{phisol2}.

The boundary conditions with $\varphi(x=0)\equiv\varphi_{B}=\pi$ (and $\varphi(x=L)=\pi+2m\pi$, $m\in\mathbb{Z}$) will be of special interest to us so we focus on these.

\subsubsection{$c> 1$}
Firstly, the amplitude is an increasing function in its argument $u+K=k^{-1}(x-x_{0})+K$ so that from the choice of signs in \eqref{phisol1} it is clear that we have a solution increasing or decreasing in $u$ from the values $\varphi_{B}=\pm\pi$.

Restricting to values of $\varphi$ with $\varphi(x=L)-\varphi(x=0)\leq 2\pi$ and choosing all positive signs we have the boundary condition at $x=0$
\bea
\varphi_{B}~~=~~\pi&=&\pi+2~{\rm am}\left(\frac{-x_{0}}{k}+K,k\right)\nonumber\\
\Ra {\rm am}\left(\frac{-x_{0}}{k}+K,k\right)&=&0\nonumber\\
\Ra x_{0}&=&kK(k)
\eea
and at $x=L$
\bea
3\pi&=&\pi+2~{\rm am}\left(\frac{L}{k},k\right)\nn\\
\Ra {\rm am}\left(\frac{L}{k},k\right)&=&\pi
\eea
which using ${\rm am}(u+2K)={\rm am}(u)+\pi$ gives us the length $L$ of the interval in terms of the elliptic modulus $k$ as
\be
L=2kK(k),\quad x_{0}=\frac{L}{2}.
\ee

\begin{figure}[!h]
\begin{center}
\includegraphics[width=0.4\textwidth]{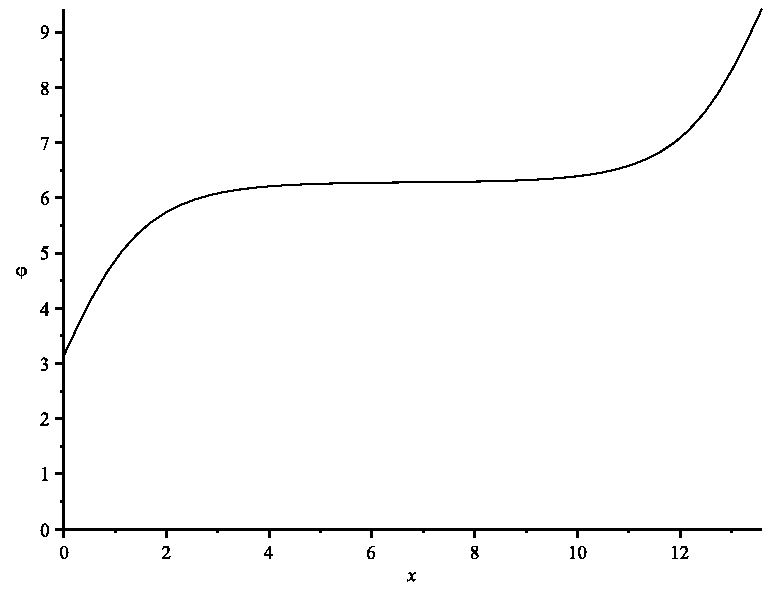}
\caption{The $c> 1$ solution plotted on $x\in[0,L]$.}
\label{Sol1plot}
\end{center}
\end{figure}

Generically, the solution increases, more rapidly at first, from $\pi$ up to $2\pi$ where the gradient $\varphi'$ decreases but remains positive, and then steepens again up to $3\pi$.  The solution is quasi-periodic with period $L$, increasing by a further $2\pi$ for each $L$ moved in $x$.  When $L\to 0$ ($k\to 0$) the solution tends to an increasingly steep, straighter line from $\pi$ to $3\pi$ while for $L\to\infty$ ($k\to 1$) the solution increases to $2\pi$ as $x\to\infty$ becoming equal to half a kink solution of the bulk theory.   The solution is plotted in Figure \ref{Sol1plot}.
\begin{figure}[!h]
\begin{center}
\includegraphics[width=0.4\textwidth]{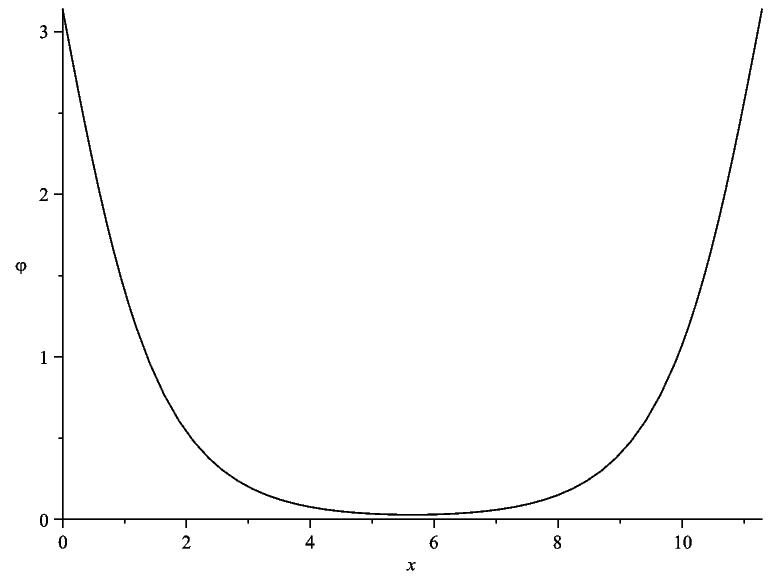}
\caption{The $|c|\leq 1$ solution plotted on $x\in[0,L]$.}
\label{Sol2plot}
\end{center}
\end{figure}
\newline

\subsubsection{$|c|\leq 1$}

In this case the solution \eqref{phisol2} is literally periodic with $0\leq\varphi\leq2\pi$ (or $2m\pi\leq\varphi\leq 2(m+1)\pi$, $m\in\mathbb{Z}$ if we'd chosen appropriately in \eqref{sGsol2}), the choice of sign reflecting the solution about $\varphi=\pi$. Taking the positive sign, at $x=0$ we have
\bea
\varphi_{B}~~=~~\pi&=&2~{\rm arccos}\left(\frac{1}{k}~{\rm sn}\left(-x_{0},\frac{1}{k}\right)\right)\nn\\
\Ra \frac{1}{k}~{\rm sn}\left(-x_{0},\frac{1}{k}\right)&=&0\nn\\
\Ra x_{0}&=&0
\eea
and if we take $x=L$ to be the point at which the solution first returns to $\varphi=\pi$ then
\bea
\pi&=&2~{\rm arccos}\left(\frac{1}{k}~{\rm sn}\left(L,\frac{1}{k}\right)\right)\nn\\
\Ra \frac{1}{k}~{\rm sn}\left(L,\frac{1}{k}\right)&=&0\nn\\
\Ra L&=&2K\left(\frac{1}{k}\right).
\eea

For generic $L(k)$ the solution decreases from $\varphi=\pi$ to some minimum at $x=\frac{L}{2}$ and then increases again up to $\varphi=\pi$.  For $L\to\infty$ ($k\to 1$) the solution approaches $\varphi=0$ as $x\to\infty$ and again becomes half a kink solution of the bulk theory in the limit.  In contrast to the $c>1$ case however when we take $k\to\infty$ we have a minimum value of L: 
\be
L_{{\rm min}}=L(k=\infty)=2K(0)=\pi.
\ee
\newline
The solution is plotted in Figure \ref{Sol2plot}.

\section{String Solutions and $\Delta-J$}

We require solutions to the O(3) sigma-model equations\footnote{In this section  we set the radius $R$ of $S^{2}$ (also of course the radius of $S^{5}$ and $AdS_{5}$) to one.}
\be
\ddot{\vec{X}}-\vec{X}''+(\dot{\vec{X}}^{2}-\vec{X}'^{2})\vec{X}=0\label{sigmaEOM}
\ee
which arise by taking the conformal and static partial gauges on the world sheet of the Polyakov string and implementing the constraint $\vec{X}^{2}=1$ by the method of Lagrange multipliers.  The squares of target space vectors are understood to mean $\vec{Y}^{2}\equiv \vec{Y}\cdot\vec{Y}\equiv Y_{i}Y^{i}=Y^{i}Y^{i}$, the index $i$ taking spacial target space values in $\mathbb{R}^{3}$, into which $S^{2}$ is embedded.  We must also satisfy the Virasoro constraints arising from taking the conformal gauge, which together with the use of the static gauge, are
\bea
\dot{\vec{X}}^{2}+\vec{X}'^{2}&=&1\label{V1}\\
\dot{\vec{X}}\cdot\vec{X}'&=&0.\label{V2}
\eea
This is achieved by mapping the solutions to sine-Gordon theory of the previous section into the string target space.  We will be particularly interested in the energy $\Delta$ of the string\footnote{Throughout this section and the next we use the symbol $\Delta$ to denote string theory energies as a direct reference to the conformal dimension of operators in $\mathcal{N}=4$ SYM to which they correspond.} and its angular momentum $J$.  In section 4 we will re-express the angular momentum and energy in a general gauge in order to make a comparison with the brane picture.

The map between sine-Gordon theory and the string sigma model is difficult to explicitly invert in general, however it is particularly easy to make the map in the case of solutions satisfying a condition due to Klose and McLoughlin\cite{Klose:2008rx} which allowed them to consider strings corresponding to various ``2 phase" sine-Gordon solutions:
\be
\partial_{x}\left(\frac{\partial_{t}\varphi}{{\rm sin}\left(\frac{\varphi}{2}\right)}\right)=0.
\ee
Clearly any static solution will do.  We therefore construct the strings corresponding to the two types of static sine-Gordon solutions following the method of Klose and McLoughlin, for which it is found that, remembering to take the radius of the sphere $R=1$ and with the coordinates
\be
\vec{X}=(r~{\rm cos}(\phi),r~{\rm sin}(\phi),\sqrt{1-r^{2}})\label{X}
\ee
the radius $r$ is given by
\be
r=-\frac{1}{\dot{\phi}}~{\rm cos}\left(\frac{\varphi}{2}\right),\quad \phi=\dot{\phi}t\label{randphi}
\ee
where
\be
\dot{\phi}=\sqrt{{\rm cos}^{2}\left(\frac{\varphi}{2}\right)+\left(\frac{\varphi'}{2}\right)^{2}}={\rm constant.}\label{phidot}
\ee
From the definition
\be
\dot{\vec{X}}^{2}={\rm cos}^{2}\left(\frac{\varphi}{2}\right)
\ee
we see that the boundary condition $\varphi_{B}=\pi$ corresponds to a stationary string endpoint, in fact stuck to the north pole of the sphere where there is taken to be a maximal ``$Z=0$'' giant graviton.  Similarly, $\varphi=\pi+2m\pi,~m\in\mathbb{Z}$ will map to a static point.  We will focus on these ``maximal'' boundary conditions.

If we work out the angular velocity $\dot{\phi}$ in each case we find that it is related to the parameter $k$, or $c$, as follows: for $c>1$, $0<k<1$ and taking $\dot{\phi}$ positive
\be
\dot{\phi}=\sqrt{{\rm sn}^{2}\left(\frac{x}{k},k\right)+\frac{1}{k^{2}}~{\rm dn}^{2}\left(\frac{x}{k},k\right)}=\frac{1}{k};\quad \phit>1
\ee
while for $|c|\leq 1$ ($c\neq -1$), $k\geq 1$, we have
\be
\dot{\phi}=\sqrt{\frac{1}{k^{2}}\left[{\rm sn}^{2}\left(x,\frac{1}{k}\right)+{\rm cn}^{2}\left(x,\frac{1}{k}\right)\right]}=\frac{1}{k};\quad 0< \phit\leq 1.
\ee
The full parameter range $-1< c$ therefore maps to the range of angular velocities $\phit> 0$, or including $c=-1$ (k strictly undefined) then $\phit\geq 0$.  Including now without harm the point $c=1,~k=1$ in both solutions we have
\be
c\geq 1 \Leftrightarrow \dot{\phi}\geq 1,\qquad |c|\leq 1 \Leftrightarrow 0\leq \dot{\phi}\leq 1\nn
\ee
so that from hereon we most usefully characterise our solutions by the value of $\phit$.

\subsection{$\dot{\phi}\geq 1$ solution}

From equations \eqref{X} and \eqref{randphi} then we have
\be
\vec{X}=\left(\frac{1}{\dot{\phi}}~{\rm cos}(\dot{\phi}t)~{\rm sn}\left(\dot{\phi}x,\frac{1}{\dot{\phi}}\right),\frac{1}{\dot{\phi}}~{\rm sin}(\dot{\phi}t)~{\rm sn}\left(\dot{\phi}x,\frac{1}{\dot{\phi}}\right),{\rm dn}\left(\dot{\phi}x,\frac{1}{\dot{\phi}}\right)\right).
\ee
It can be checked that this solution satisfies the string's sigma model equations of motion \eqref{sigmaEOM}, both Virasoro constraints \eqref{V1} and \eqref{V2}, and returns the correct form of $\varphi(x)$ using the sine-Gordon map.

\begin{figure}
\begin{center}
\includegraphics[width=0.4\textwidth]{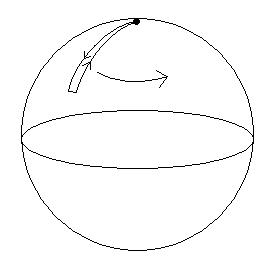}
\hspace{1.5in}\parbox{5in}{\caption{The $\dot{\phi}>1$ string depicted with halves of the string artificially separated for clarity.  For $\dot{\phi}=1$ the cusp point reaches the equator while for $\dot{\phi}\to\infty$ the string retreats to the north pole where it vanishes.\label{String01}}}
\end{center}
\end{figure}

The solution is a folded `half a spinning string' with both endpoints attached to the north pole.  For $\dot{\phi}\to 1$ the cusp point reaches to the equator, while for $\dot{\phi}\to\infty$ the string shortens toward a point at the pole.  The cusp point always maintains the condition
\be
r\dot{\phi}=1,\quad\Ra\quad {\rm cos}\left(\frac{\varphi}{2}\right)=-1
\ee
i.e. we always have a point at which $\varphi=2\pi$, which is true of the sine-Gordon solution in this parameter range.  The string is depicted in Figure \ref{String01}.
\newline

We now compute the angular momentum of the string as
\be
J=\frac{\sqrt{\lambda}}{2\pi}\int_{0}^{L}~{\rm d}x~\vec{X}\wedge\dot{\vec{X}}=\frac{\sqrt{\lambda}}{2\pi}\frac{1}{\dot{\phi}}\int_{x=0}^{x=L}~{\rm d}u~\vec{X}\wedge\dot{\vec{X}},\quad u=\dot{\phi}x.
\ee
We have
\be
\vec{X}\wedge\dot{\vec{X}}=\left(
\begin{array}{c}
{\rm cos}(\dot{\phi}t)~{\rm sn}(u,k)~{\rm dn}(u,k)\\-{\rm sin}(\dot{\phi}t)~{\rm sn}(u,k)~{\rm dn}(u,k)\\~\frac{1}{\dot{\phi}}{\rm sn}^{2}(u,k)
\end{array}\right).
\ee
Using the integrals
\be
\int~{\rm d}u~{\rm sn}(u,k){\rm dn}(u,k)=-{\rm cn}(u,k),\qquad\int~{\rm d}u~{\rm sn}^{2}(u,k)=\frac{u-E(u,k)}{k^{2}}\nn
\ee
where $E(u,k)$ is the incomplete elliptic integral of the second kind and $k=\frac{1}{\dot{\phi}}$ we find
\be
\vec{J}=\frac{\sqrt{\lambda}}{2\pi}\left(\frac{2~{\rm cos}(\dot{\phi}t)}{\dot{\phi}},~\frac{2~{\rm sin}(\dot{\phi}t)}{\dot{\phi}},~2(K(k)-E(k))\right)
\ee
where $E(k)$ is the \emph{complete} elliptic integral of the second kind and $K(k)$ that of the first.  We will be interested in the $z$-component of the angular momentum, thus we take
\be
J_{1}\equiv J_{z}=\frac{\sqrt{\lambda}}{\pi}(K(k)-E(k)),\qquad k=\frac{1}{\phit}.\label{J1}
\ee

The energy $\Delta$ of the string in the conformal gauge is just the length of the string in $x$ multiplied by the string tension, or the length of the sine-Gordon interval times the tension.  Hence,
\bea
\Delta_{1}&=&\frac{\sqrt{\lambda}}{2\pi}~L\nn\\
&=&\frac{\sqrt{\lambda}}{\pi}~kK(k),\qquad k=\frac{1}{\phit}.\label{Delta1}
\eea

\subsection{$0\leq \dot{\phi}\leq 1$ solution}

This time equations \eqref{X} and \eqref{randphi} give us
\be
\vec{X}=\left({\rm cos}(\dot{\phi}t)~{\rm sn}(x,\dot{\phi}),~{\rm sin}(\dot{\phi}t)~{\rm sn}(x,\dot{\phi}),-{\rm cn}(x,\dot{\phi})\right).
\ee
Again, it can be checked that this solution satisfies the string's sigma model equations of motion \eqref{sigmaEOM}, both Virasoro constraints \eqref{V1} and \eqref{V2}, and returns the correct form of $\varphi(x)$ using the sine-Gordon map.

\begin{figure}[!h]
\begin{center}
\includegraphics[width=0.4\textwidth]{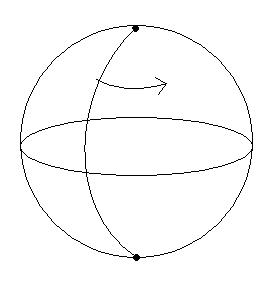}
\hspace{1.5in}\parbox{5in}{\caption{The $0\leq \dot{\phi}\leq 1$ string.  The string is geometrically indistinguishable for different valid values of $\dot{\phi}$, apart from moving at different angular velocities.\label{String02}}}
\end{center}
\end{figure}

The solution is qualitatively different from the first, being a rotating, \emph{stretched} string between the north and south poles of the sphere.  For $\dot{\phi}\to 1$ it can in fact be seen that the two solutions become identical except that for $\dot{\phi}>1$ half the string is folded back onto the same hemisphere while for $0\leq \dot{\phi}\leq 1$ the string continues through the equator to the south pole.  As $\dot{\phi}\to 0$ we obtain just the stretched string on $S^{2}$.  We have no cusp point; the corresponding sine-Gordon solution never passes through $\varphi=2m\pi,~m\in\mathbb{Z}$.  The solution is depicted in Figure \ref{String02}.
\newline

Computing the angular momentum as above,
\be
\vec{X}\wedge\dot{\vec{X}}=\left(
\begin{array}{c}
\dot{\phi}~{\rm cos}(\dot{\phi}t)~{\rm cn}(u,\dot{\phi})~{\rm sn}(u,\dot{\phi})\\-\dot{\phi}~{\rm sin}(\dot{\phi}t)~{\rm cn}(u,\dot{\phi})~{\rm sn}(u,\dot{\phi})\\~\dot{\phi}~{\rm sn}^{2}(u,\dot{\phi})
\end{array}\right)
\ee
and we further need the integral
\be
\int~{\rm d}u~{\rm cn(u,m})~{\rm sn}(u,m)=-\frac{{\rm dn}(u,m)}{m^{2}},\quad u=x,~m=\dot{\phi}\nn
\ee
giving
\be
\vec{J}=\frac{\sqrt{\lambda}}{2\pi}\left(0~,~0,\frac{2(K(\dot{\phi})-E(\dot{\phi}))}{\dot{\phi}}\right).
\ee
The only non-zero component this time is the $z$-component:
\be
J_{2}\equiv J_{z}=\frac{\sqrt{\lambda}}{\pi}\frac{(K(\dot{\phi})-E(\dot{\phi}))}{\dot{\phi}}.\label{J2}
\ee
Again, the energy is just proportional to the length of the string:
\bea
\Delta_{2}&=&\frac{\sqrt{\lambda}}{2\pi}~L\nn\\
&=&\frac{\sqrt{\lambda}}{\pi}~K(\dot{\phi}).\label{Delta2}
\eea

\subsection{$\Delta-J$ and finite $J$ corrections}

\begin{figure}[!h]
\begin{center}
\includegraphics[width=0.5\textwidth]{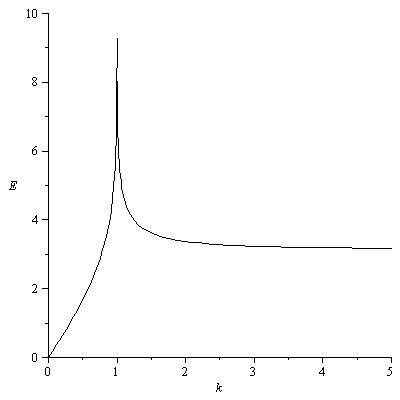}
\hspace{1.5in}\parbox{5in}{\caption{The energy in units of $\sqrt{\lambda}/2\pi$ of both solutions plotted against $k=\dot{\phi}^{-1}$. The energy of the $0\leq \dot{\phi}\leq 1$ solution tends to that of the static, stretched string as $\dot{\phi}\to 0$, or $k\to\infty$.\label{E_vs_k}}}
\end{center}
\end{figure}

\begin{figure}[!h]
\begin{center}
\includegraphics[width=0.5\textwidth]{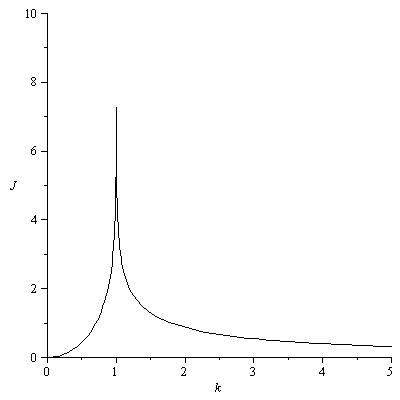}
\hspace{1.5in}\parbox{5in}{\caption{Angular momentum in units of $\sqrt{\lambda}/2\pi$ plotted against $k=\dot{\phi}^{-1}$. The $\dot{\phi}\geq 1$ solution lies to the left of the divergence and the $0\leq \dot{\phi}\leq 1$ solution lies to the right.  For both $\dot{\phi}\to 0$ ($k\to \infty$) and $\dot{\phi}\to \infty$ ($k\to 0$) the angular momentum vanishes, in the former case because the stretched string is static, in the latter case because the string itself vanishes.\label{J_vs_k}}}
\end{center}
\end{figure}

Both the energy and angular momentum of these solutions are divergent when $\dot{\phi}\to 1$ ($k\to 1$), as can be seen in Figures \ref{E_vs_k} and \ref{J_vs_k}.  Both solutions possess a limit in which the angular momentum vanishes, but only the $\dot{\phi}>1$ solution has a vanishing energy (as $\dot{\phi}\to\infty$, or $k\to 0$) - the $0\leq \dot{\phi}\leq 1$ solution has a minimum energy which is that of a stretched string.  In the previous section we remarked that the $|c|\leq 1$ sine-Gordon solution possessed a minimum interval length given by
\be
L_{{\rm min}}=2K(0)=\pi
\ee
which in turns gives us a minimum energy
\be
\Delta_{2,{\rm min}}=\frac{\sqrt{\lambda}}{2\pi}~\pi=\frac{\sqrt{\lambda}}{2}
\ee
which is indeed the energy of a string of tension $\sqrt{\lambda}/2\pi$ stretched to a target space length of $\pi$.

\begin{figure}[!h]
\begin{center}
\includegraphics[width=0.5\textwidth]{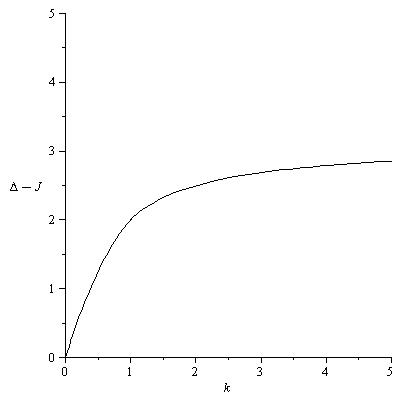}
\hspace{1.5in}\parbox{5in}{\caption{Plot of $\Delta-J$ in units of $\sqrt{\lambda}/2\pi$ for both solutions.  Both are finite for all $k$, or $\dot{\phi}$, and in particular we move smoothly between the solutions as we go through $\dot{\phi}=k=1$.\label{EJ_vs_k}}}
\end{center}
\end{figure}

Figure \ref{EJ_vs_k} plots $\Delta-J$ for both solutions.  The quantity is always finite, vanishing at $\dot{\phi}\to \infty$ ($k\to 0$) and tending to a constant value as $\dot{\phi}\to 0$ ($k\to\infty$).  In the limit $\dot{\phi}\to 1$ both solutions tend to the same configuration (bar the change of hemisphere of one half of the string) which is that with
\be
\Delta-J=\frac{\sqrt{\lambda}}{\pi}
\ee
studied previously.

We can examine the leading order finite $J$ corrections to both of these solutions by expanding $\Delta$ and $J$ around $\dot{\phi}=k=1$, with $\epsilon\equiv\sqrt{1-k^{2}}$, finding $\epsilon(J)$ and then resubstituting this back into our expressions for $\Delta-J$.  Using the following expansions
\bea
K(\epsilon)&\approx&-{\rm ln}\left(\frac{\epsilon}{4}\right)+\frac{\epsilon^{2}}{4}\left(-{\rm ln}\left(\frac{\epsilon}{4}\right)-1\right)\\
E(\epsilon)&\approx&1+\frac{\epsilon^{2}}{2}\left(-{\rm ln}\left(\frac{\epsilon}{4}\right)-\frac{1}{2}\right)
\eea
we have from equation \eqref{J1}
\bea
\frac{2\pi}{\sqrt{\lambda}}~J_{1}&=&2(K(k)-E(k))\nn\\
&\approx&2\left(-{\rm ln}\left(\frac{\epsilon}{4}\right)-1\right)\nn\\
\Ra -\frac{2\pi}{\sqrt{\lambda}}~J_{1}&\approx&{\rm ln}\left(\epsilon^{2}\frac{e^{2}}{16}\right)\nn\\
\eea
giving $\epsilon(J)$ as
\be
\epsilon^{2}=\frac{16}{e^{2}}e^{-\frac{2\pi}{\sqrt{\lambda}}J}.\label{epsilonJ1}
\ee
From equations \eqref{J1} and \eqref{Delta1} we have
\bea
\frac{2\pi}{\sqrt{\lambda}}(\Delta_{1}-J_{1})&=&2\left[(k-1)K(k)+E(k)\right]\nn\\
&\approx&2\left(1-\frac{\epsilon^{2}}{2}\right)\left(-{\rm ln}\left(\frac{\epsilon}{4}\right)+\frac{\epsilon^{2}}{4}\left(-{\rm ln}\left(\frac{\epsilon}{4}\right)-1\right)\right)\nn\\
&&+2+\epsilon^{2}\left(-{\rm ln}\left(\frac{\epsilon}{4}\right)-\frac{1}{2}\right)\nn\\
&\approx&2\left[1-\frac{\epsilon^{2}}{4}\right]\nn
\eea
or
\be
\Delta_{1}-J_{1}\approx\frac{\sqrt{\lambda}}{\pi}\left[1-\frac{4}{e^{2}}e^{-\frac{2\pi}{\sqrt{\lambda}}J}\right]\label{DeltaJ1}
\ee
matching the correction \eqref{finiteJBak} found by Bak\cite{Bak} for the very same string configuration, which as remarked in the introduction is seen to be exactly two coincident halves of a normal giant magnon with $p=\pi$.

Note that if we perform the same analysis for the $0\leq\dot{\phi}\leq 1$ solution then, as seen from Figure \ref{EJ_vs_k}, the correction should be equal in magnitude and opposite in sign, $\Delta-J$ being continuous as we move between the two solutions.

This time then the angular momentum is
\be
J_{2}=\frac{\sqrt{\lambda}}{2\pi}\frac{2(K(\dot{\phi})-E(\dot{\phi}))}{\dot{\phi}}
\ee
where $\dot{\phi}$ is playing the role of the elliptic modulus.  Hence we expand in $\varepsilon=\sqrt{1-\dot{\phi}^{2}}$ instead and at leading order find
\be
\frac{2\pi}{\sqrt{\lambda}}J_{2}\approx2\left(-{\rm ln}\left(\frac{\varepsilon}{4}\right)-1\right)
\ee
which similarly to \eqref{epsilonJ1} gives
\be
\varepsilon^{2}=\frac{16}{e^{2}}e^{-\frac{2\pi}{\sqrt{\lambda}}J}.\label{epsilonJ2}
\ee
Then from equations \eqref{J2} and \eqref{Delta2} we have
\bea
\frac{2\pi}{\sqrt{\lambda}}(\Delta_{2}-J_{2})&=&2\left(K(\dot{\phi})-\frac{K(\dot{\phi})-E(\dot{\phi})}{\dot{\phi}}\right)\nn\\
&=&2\left[1-\frac{\varepsilon^{2}}{2}{\rm ln}\left(\frac{\varepsilon}{4}\right)-\frac{\varepsilon^{2}}{4}\right]+O\left(\varepsilon^{4}{\rm ln}(\varepsilon)\right)\nn\\
&\approx&2\left[1-\frac{\varepsilon^{2}}{2}\left(-\frac{2\pi}{\sqrt{\lambda}}J-1\right)-\frac{\varepsilon^{2}}{4}\right]+O\left(\varepsilon^{4}{\rm ln}(\varepsilon)\right)
\eea
becoming
\be
\Delta_{2}-J_{2}\approx\frac{\sqrt{\lambda}}{\pi}\left[1+\frac{4}{e^{2}}e^{-\frac{2\pi}{\sqrt{\lambda}}J}\right]+e^{-\frac{2\pi}{\sqrt{\lambda}}J}J.
\ee
Unlike in the $\dot{\phi}>1$ case we have a term not proportional to $\sqrt{\lambda}$.  With the interpretation that the string solution is valid at large $\lambda$ then the `extra' term becomes negligible, such that
\be
\Delta_{2}-J_{2}\approx\frac{\sqrt{\lambda}}{\pi}\left[1+\frac{4}{e^{2}}e^{-\frac{2\pi}{\sqrt{\lambda}}J}\right]\label{DeltaJ2}
\ee
which as predicted is the same as equation \eqref{DeltaJ1} but for a difference of sign in the correction term.

\section{D3-Brane BIon Solutions}

In this section we will examine the world volume theory of the giant graviton wrapping an $S^{3}\subset S^{5}$ in order to rediscover the range of behaviour exhibited by the finite $J$ boundary giant magnons of the previous section.  The strings appear in an analogous manner to that of the BIon spikes of D3-branes in flat space\cite{Callan:1997kz,Gibbons:1997xz} requiring both scalar and gauge fields to be excited.  Related solutions for the giant graviton at $\phit=1$ were constructed in \cite{Hirano,Sadri:2003mx} while here we allow any $\phit\geq 0$.

We use the static gauge where the world volume time $\tau$ is chosen to coincide with the target space time coordinate $t$ that appears as the global time variable of the $AdS$ part of $AdS_{5}\times S^{5}$. With the abbreviations $\rho^{2}\equiv 1-r^{2}$, $s_{4}\equiv {\rm sin}\sigma_{4}$ and $s\equiv {\rm sin}\sigma$, we may write the metric on $\mathbb{R}\times S^{5}$ as
\be
{\rm d}s^{2}=R^{2}\left\{-{\rm d}t^{2}+\frac{{\rm d}r^{2}}{\rho^{2}}+r^{2}{\rm d}\phi^{2}+\rho^{2}({\rm d}\sigma^{2}+s^{2}{\rm d}\sigma_{4}^{2}+s^{2}s_{4}^{2}{\rm d}\sigma_{5}^{2})\right\}.
\ee
The D3-brane is taken to lie at the centre of $AdS_{5}$ so that the world volume scalars associated with these directions are unexcited.  We will mainly be concerned with the maximal giant graviton that wraps an $S^{3}$ entirely transverse to the string coordinates (a $Z=0$ giant graviton).
\newline

We first examine the Nambu string for later comparison with the BIons, using the same coordinate system.  

\subsection{Nambu string}

We are going to produce Born-Infeld configurations that have a target space interpretation of fundamental strings attaching to the giant graviton.  In order that we are able to identify the appearance of the string correctly we first find expressions for the angular momentum and energy of the string in a general gauge.

The solutions we are after live on an $S^{2}$ of radius $R$ which we choose to be coordinated by $\{r,\phi\}$, and are radially extended in $r$, so that denoting with a prime derivatives with respect to the world sheet spacial coordinate $x$ we have $r'\neq 0$ and $\phi'=0$. The strings move rigidly with $\dot{r}=\ddot{\phi}=0$.  With induced metric $g_{ab}$ the Nambu-Goto Lagrangian density is
\bea
\mathcal{L}_{{\rm NG}}&=&-\frac{1}{2\pi l_{s}^{2}}\sqrt{-{\rm det}g_{ab}}\nn\\
&=&-\frac{\sqrt{\lambda}}{2\pi}\Lambda r',
\eea
where we defined $\Lambda\equiv \sqrt{\frac{1-\phit r^{2}}{1-r^{2}}}$.  The angular momentum density is
\be
j_{{\rm s}}=\slambda\frac{r^{2}r'}{\rho^{2}}\frac{\phit}{\Lambda}.\label{js}
\ee
Given the dependence here only on $r$, $\phit$ and a single power of $r'$ we may integrate up immediately with respect to $r$.  In light of the above results we should find two sets of solutions depending upon whether $0\leq\phit\leq 1$ or $\phit \geq 1$.  If for $\phit\geq 1$ we take the upper and lower limits of $r$ to be $r=\phit^{-1}$ and $r=0$ respectively, and for $0\leq \phit\leq 1$ we take $r=1$ and $r=0$ (so that in both cases we shall obtain half of one of the strings above) then we find
\bea
J_{{\rm s}}&=&-\slambda\int^{r_{{\rm U}}}_{r_{{\rm L}}}\frac{r^{2}}{\rho^{2}}\frac{\phit}{\Lambda}{\rm d}r\label{Js1}\\
&=&\Bigg\{\begin{array}{cc}
-\slambda\left(E\left(\frac{1}{\phit}\right)-K\left(\frac{1}{\phit}\right)\right),& \mbox{if }\phit\geq 1\\
-\slambda\frac{1}{\phit}\left(E(\phit)-K(\phit)\right),& \qquad\mbox{if }0\leq \phit\leq 1
\end{array}\label{Js2}
\eea
which is precisely as expected from the sine-Gordon derived results, without even requiring the solutions themselves.  The overall minus signs reflect the fact that the world sheet spacial variable runs from the end of the string at or closest to the equator, or the \emph{upper} limit of $r$, to the point with $r=0$.

The Nambu string energy density $h_{{\rm s}}$ is given by
\be
h_{{\rm s}}=\slambda\frac{r'}{\rho^{2}}\frac{1}{\Lambda}\label{hs}
\ee 
which when integrated for the string energy $\Delta_{{\rm s}}$ returns the results obtained above for any $\phit\geq 0$ and therefore returns the same leading order correction to the quantity $\Delta_{{\rm s}}-J_{{\rm s}}$:
\be
\Delta_{{\rm s}}-J_{{\rm s}}\approx\slambda\left\{1\pm\frac{4}{e^{2}}e^{-\frac{2\pi}{\sqrt{\lambda}}J_{{\rm s}}}\right\}\label{EJs}
\ee
where the plus sign is for $\phit\geq 1$ and the minus sign is for $0\leq\phit\leq 1$.  Note that this is in fact half of the $\Delta-J$ calculated in the previous section simply because we have integrated over a single half of the full string.

\subsection{D3 action and ansatz}

Next we turn to the Born-Infeld theory itself.  The brane action, having kinetic and potential parts given by the Born-Infeld action and the Chern-Simons coupling respectively, is
\be
S=-T_{{\rm D}3}\left[\int_{\Omega_{{\rm D}3}}\sqrt{-{\rm det}(g_{ab}+2\pi l_{s}^{2}F_{ab})}+\int_{\Omega_{{\rm D}3}}~C_{4}\right]
\ee
where $g_{ab}$ is the metric induced on the world volume of the D3 brane embedded in $S^{5}$ so the indices $a,b$ take one time and 3 spacial values; $F_{ab}$ is the field strength tensor for the electromagnetic fields living on the world volume.

We choose the embedding of the D3 brane to be given simply by the coordinates \{$\sigma,~\sigma_{4},~\sigma_{5}$\} so that the coordinates \{$r,~\phi$\} are transverse to the brane and equal to two of the scalar fields living thereon (there will be 4 more that will not concern us).

We can simplify the explicit form of the action if we at this point make the ansatz that the scalar fields depend only upon the coordinate $\sigma$, with $\dot{r}=0$ and $\dot{\phi}\neq 0$ but constant.  Brane solutions will therefore be symmetric with respect to rotations by $\sigma_{4}$ and $\sigma_{5}$.  We will now denote with a prime the derivative with respect to $\sigma$
and take a purely electric $F_{ab}$ with only $F_{\tau\sigma}=-F_{\sigma \tau}\neq 0$.

The potential term is an integral over the world volume of the brane $\Omega_{{\rm D}3}$ of the background 4-form potential $C_{4}$.  This can be rewritten in terms of the 5-form field strength by
\be
S_{{\rm CS}}=\int_{\Omega_{{\rm D}3}}C_{4}=\int_{\Sigma}F_{5}
\ee
where $F_{5}={\rm d}C_{4}$ and $\Sigma$ is now the 5-manifold whose boundary is the 4 dimensional hypersurface swept out by the D3 brane.  Since the background flux has a constant density over the $S^{5}$, $F=B{\rm dvol}_{5}$ so that
\bea
S_{{\rm CS}}&=&B~{\rm vol}(\Sigma)\nn\\
&=&B \int_{\Sigma} R^{5}r\rho^{2}s^{2}s_{4}{\rm d}r{\rm d}\phi{\rm d}\sigma{\rm d}\sigma_{4}{\rm d}\sigma_{5}.\nn
\eea
We may immediately integrate over $\sigma_{4}$ and $\sigma_{5}$ then convert from an integral over $r$, at fixed $\sigma$, to $\rho$ which goes from 0 up to the dimensionless radius of the brane $\rho(\sigma)$, and finally use ${\rm d}\phi=\dot{\phi}~{\rm d}t$ to get a Lagrangian density for the potential of
\be
\mathcal{L}_{CS}=\pi BR^{5}~{\rm sin}^{2}\sigma~\rho^{4}\dot{\phi}.
\ee
$B$ may be re-expressed using the flux quantisation condition for the 5-from on $S_{5}$, $BR^{5}\Omega_{5}=2\pi N$, while the  D3-brane tension is
\bea
T_{p=3}&=&\frac{2\pi}{g_{s}(2\pi l_{s})^{p+1}}\Big|_{p=3}\\
\Ra ~~T_{3}&=&\frac{1}{g_{s}(2\pi)^{3}l_{s}^{4}}\nn\\
&=&\frac{N}{R^{4}}\frac{1}{\Omega_{3}},\qquad \Omega_{3}=2\pi^{2}
\eea
where we used the relation between the forms of the fixed 't Hooft coupling $\lambda$,
\be
\quad\lambda=\frac{R^{4}}{l_{s}^{4}}=4\pi g_{s}N.
\ee
Hence we can write the potential term as
\be
\mathcal{L_{{\rm CS}}}=T_{3}\Omega_{2}R^{4}~{\rm sin}^{2}\sigma~(1-r^{2})^{2}\dot{\phi}.
\ee

Putting these together the Lagrangian density in $\sigma$ is
\be
\mathcal{L}=-T\int~{\rm d}\sigma{\rm d}\tau~s^{2}\left[\rho^{2}\sqrt{D}-\rho^{4}\dot{\phi}\right]
\ee
with effective tension $T=4\pi R^{4}T_{3}$ and
\be
D=r'^{2}\Lambda^{2}+r^{2}\phi'^{2}-\tilde{F}^{2}+\rho^{4}\Lambda^{2}
\ee
with rescaled electric field $\tilde{F}=\left(\frac{2\pi l_{s}^{2}}{R^{2}}\right)F_{\tau\sigma}$.

Turning to the equations of motion we encounter only total derivatives with respect to $\sigma$ due to the time independence of our ansatz, thus for example we have
\be
\frac{\partial}{\partial\tau}\frac{\partial\mathcal{L}}{\partial \dot{\phi}}=-Ts^{2}~\frac{\partial}{\partial\tau}\left(\frac{1}{2}\frac{1}{\sqrt{D}}\frac{\partial D}{\partial \dot{\phi}}-\rho^{4}\right)=0
\ee
as everything is independent of time except $\phi$ itself, which does not appear.  Then the $\phi$ equation of motion becomes
\be
\phi:\qquad\frac{{\rm d}}{{\rm d}\sigma}\left(\frac{s^{2}\rho^{2}r^{2}\phi'}{\sqrt{D}}\right)=0.\label{phiEOM}
\ee
As we have already covered we would like to study a radial configuration so that $\phi'=0$ and the $\phi$ equation is therefore satisfied automatically by these configurations.  The equation of motion for the gauge component $A_{\sigma}$ is similarly trivial while for the $A_{\tau}$ component we have
\be
A_{\tau}:\qquad\frac{{\rm d}}{{\rm d}\sigma}\left(\frac{s^{2}\rho^{2}\tilde{F}}{\sqrt{D}}\right)=0\label{AEOM}
\ee
and for the $r$ equation of motion we have
\be
r:\qquad\frac{{\rm d}}{{\rm d}\sigma}\left(\frac{s^{2}\rho^{2}r'\Lambda^{2}}{\sqrt{D}}\right)=s^{2}\frac{\partial}{\partial r}\left(\rho^{2}\sqrt{D}-\rho^{4}\phit\right).\label{rEOM}
\ee
\newline

The $A_{\tau}$ equation can be integrated once to give a constant in $\sigma$:
\be
\frac{s^{2}\rho^{2}\tilde{F}}{\sqrt{D}}=\kappa.
\ee
The constant is fixed by the electric flux quantisation condition\footnote{The number of F-strings attaching to the brane is therefore given by the integer $k$.} $\Pi_{A}=k\in\mathbb{Z}$;
\bea
\Pi_{A}&=&\frac{\partial \mathcal{L}}{\partial \dot{A_{\tau}}}=\frac{\partial \mathcal{L}}{\partial F_{\tau\sigma}}\frac{\partial F_{\tau\sigma}}{\partial A_{\tau}'}=-\frac{\partial \mathcal{L}}{\partial F_{\tau\sigma}}\nn\\
&=&-T\frac{s^{2}\rho^{2}}{\sqrt{D}}\cdot\left(\frac{2\pi l_{s}^{2}}{R^{2}}\right)^{2}F_{\tau\sigma}
\eea
and then using the $A_{\tau}$ equation of motion, \eqref{AEOM}, we can write
\be
T\kappa\left(\frac{2\pi l_{s}^{2}}{R^{2}}\right)=k,~~\Rightarrow~~\kappa=\frac{1}{T}\left(\frac{R^{2}}{2\pi l_{s}^{2}}\right)k=\frac{\sqrt{\lambda}}{4N}k\label{kappa}
\ee
recovering the form of the constant introduced in \cite{Hirano}.

The angular momentum density is found to be
\be
\Pi_{\phi}=Ts^{2}\left[\frac{r^{2}\phit}{\sqrt{D}}\left(r'^{2}+\rho^{4}\right)+\rho^{4}\right]\label{Piphi}
\ee
and the Hamiltonian density is
\bea
\mathcal{H}&=&\Pi_{\phi}\phit+\Pi_{A}F_{\tau\sigma}-\mathcal{L}\label{H1}\\
&=&\frac{Ts^{2}}{\sqrt{D}}\left(r'^{2}+\rho^{4}\right).\label{H2}
\eea
By rearranging the $A_{\tau}$ equation \eqref{AEOM} we can find a general form of the field strength $\F$ in terms of the derivative $r'$:
\be
\F=\pm\kappa\Lambda\sqrt{\frac{r'^{2}+\rho^{4}}{s^{4}\rho^{4}+\kappa^{2}}}.\label{F}
\ee

Before proceeding to the case of general $\phit$ we specialise to $\phit=1$ to recover known results.\footnote{Most of the results in the following subsection are to be found in a slightly different form in \cite{Hirano}.}

\subsection{$\phit=1$}

\begin{figure}[h]
\begin{center}
\includegraphics[width=0.5\textwidth]{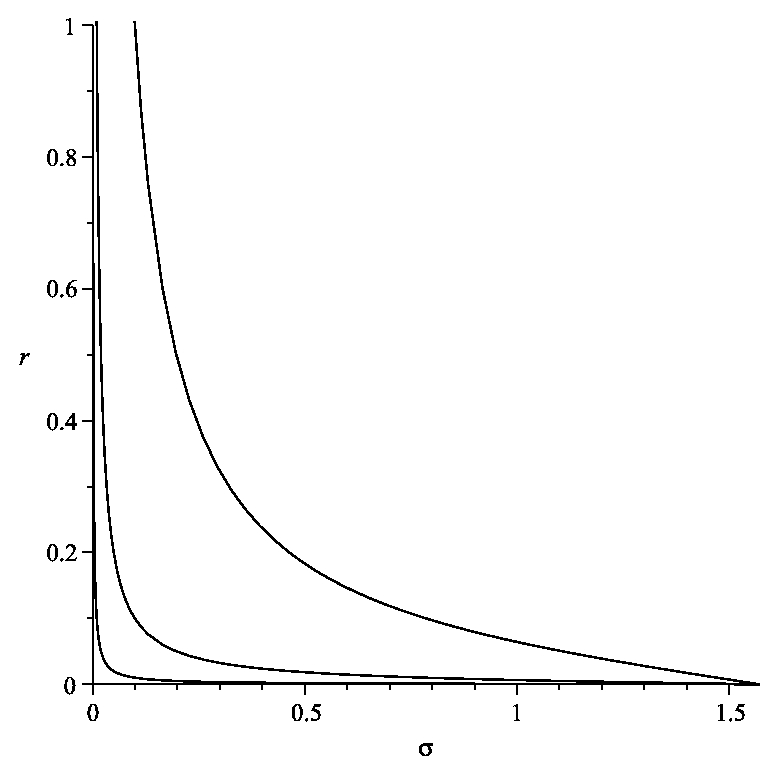}
\hspace{1.5in}\parbox{5in}{\caption{The $\phit=1$ solution with $c=0$ and positive sign taken plotted for three values of $\kappa$: $\kappa=0.1$, $\kappa=0.01$, $\kappa=0.001$.  The solution has boundary conditions $r\left(\frac{\pi}{2}\right)=0$ and $r(\sigma_{0})=1$.  As $\kappa\to 0$ the solution tends to a constant for most of $\sigma$ with a sudden spike as we approach $\sigma_{0}\to 0$.\label{plotphit1soln}}}
\end{center}
\end{figure}

One way to recover the $\phit=1$ solution is to use the $A_{\tau}$ equation \eqref{AEOM} to write $\sqrt{D}=\frac{s^{2}\rho^{2}\tilde{F}}{{\kappa}}$ and set $\phit=1$ where ever else it appears
so that the $r$ equation becomes
\be
\frac{{\rm d}}{{\rm d}\sigma}\left(\frac{r'}{\F}\right)=-2r\rho^{2}\left[\left(\frac{s^{2}\F}{\kappa}\right)^{2}-1\right]^{2}.
\ee
This can be satisfied if
\be
\F=r'\quad {\rm and}\quad\frac{s^{2}\F}{\kappa}=\pm1\quad\Ra\quad r'=\pm\frac{\kappa}{s^{2}}.\label{phit1eqn}
\ee
To check for consistency with the $A_{\tau}$ equation we substitute $\kappa=s^{2}r'$ (and $\phit=1$) into the general form of $\F$ given by equation \eqref{F} and this indeed returns us $\F=r'$.  Solving equation \eqref{phit1eqn} gives the solution
\be
r=c\mp\kappa{\rm cot}\sigma.\label{phit1soln}
\ee
This is essentially the solution presented in \cite{Hirano} (and similarly in \cite{Sadri:2003mx}) except that we have solved for $r$ as opposed to one of the Cartesian coordinates \{$x_{1}, x_{2}$\} satisfying $x_{1}^{2}+x_{2}^{2}=r^{2}$, and we have set $\phi'=0$.  There $x_{1}$ was taken to be constant, giving the familiar geometry of the giant magnon as a chord within the disc formed when  projecting the $S^{2}$ onto the $\{x_{1},x_{2}\}$ plane.  Our analogous solution effectively has $x_{1}=0$ so that the string/BIon protrudes radially from the giant graviton situated at $r=0$.  For the constant\footnote{The constant $c$ appearing here is not the constant introduced in section 2.} $c=0$ the body of the D-brane sits at $r=0$ corresponding to a maximal giant graviton.  The solution is plotted in Figure \ref{plotphit1soln}.

While the solution of \cite{Hirano} ran from negative to positive $x_{2}(\sigma)$, $r$ can only be positive or zero.  However, by taking the appropriate sign we can maintain $r\geq 0$.  In this way the solution is well defined on $\sigma\in[\sigma_{0},\pi-\sigma_{0}]$ and describes two spikes at either end of the range of $\sigma$.  If the range of $\sigma$ is $\sigma_{0}\leq\sigma\leq\pi-\sigma_{0}$ (and taking $c=0$) then we have that $r(\sigma_{0})=r(\pi-\sigma_{0})=1$.  We could choose the spikes to emerge in opposite directions, leading to precisely the previous solution with $x_{1}=0$ and corresponding to the giant magnon string solution with world sheet momentum $p=\pi$ (alternately a sine-Gordon kink with velocity $v=1$).  Or we could take both spikes to emerge in the same direction leading to two new solutions that are only physical for the strings that pass over the poles of the sphere, i.e. $r=0$; they are the `boundary giant magnons', one of which is described in \cite{Bak}.  If $x_{3}$ is the third Cartesian coordinate on $S^{2}$ satisfying $r^{2}+x_{3}^{2}=1$ then we may choose either $0\leq \pm x_{3}\leq 1$, placing both spikes on just one of the hemispheres of $S^{2}$, or choose $-1\leq x_{3}\leq 1$ so that the string passes from pole to pole through the equator.

From equation \eqref{kappa} we have that $\kappa=\frac{\pi}{\sqrt{\lambda}}g_{{\rm s}}$ so that with $\lambda$ fixed $\kappa$ controls the string coupling.  The Nambu string description in which the giant magnons live requires vanishing string coupling and hence $\kappa\to 0$. As $\kappa\to 0$ the spikes on the brane become more pronounced, being concentrated into the points $\sigma_{0}$ and $\pi-\sigma_{0}$, while in between these points the radius of the brane (that is, the radius $R\rho$ of the $S^{2}$ at each point in $\sigma$) tends to a constant.  In the limit then, the picture is of a spherical giant graviton with infinitesimally thin strings attached to its poles.  In the same limit $\sigma_{0}$, for which $r(\sigma_{0})=r(\pi-\sigma_{0})=1$, tends to zero.

Taking non-zero values of the constant $c$ was not discussed in \cite{Hirano} but it would appear to describe boundary giant magnons attached to non-maximal $Z=0$ giant gravitons, at least for small values of $c$ where we can trust the brane description of the graviton state. In sine-Gordon theory these solutions are described by taking boundary values of $\varphi\neq \pi$.  The non-maximal but, as we shall see below, still BPS giant graviton now rotates in the $\{x_{1},x_{2}\}$ plane maintaining $\phit=1$.

This behaviour covers that which we have found for boundary giant magnons in the previous section, described by sine-Gordon theory on the interval at $\phit=1$, or $L\to\infty$.  Below we shall find approximate solutions to the brane equations that reproduce the behaviour of the two types of boundary giant magnons at $\phit\neq 1$.
\newline

At $\phit=1$ the factor $\sqrt{D}=\rho^{2}$ and so the total angular momentum density given by \eqref{Piphi} contains a term $\propto r'$ and two that are not.  If we take the $r'$ term as the string's contribution to the angular momentum density, and take a single string ($k=1$), then integrating over one of the spikes on $\sigma\in[\sigma_{0},\frac{\pi}{2}]$ we get the divergent quantity
\bea
J|_{\phit=1}&=&T\int_{\sigma_{0}}^{\frac{\pi}{2}}\frac{s^{2}r^{2}r'^{2}}{\rho^{2}}{\rm d}\sigma\nn\\
&=&T\kappa\int_{\sigma_{0}}^{\frac{\pi}{2}}\frac{r^{2}r'}{\rho^{2}}{\rm d}\sigma\nn\\
&=&-\slambda\int_{0}^{1}\frac{r^{2}}{\rho^{2}}{\rm d}r=J_{{\rm s}}|_{\phit=1}.
\eea
In the last step we used
\be
r'=\frac{\kappa}{s^{2}},\quad\kappa=\frac{\sqrt{\lambda}}{4N},\quad T=\frac{2}{\pi}N,\quad\Ra T\kappa=\slambda
\ee 
which recovers the tension of the fundamental string from the brane picture. This angular momentum is equal to that of the Nambu string at $\phit=1$, as seen by substituting $\phit=1$ into equations \eqref{js} and \eqref{Js1}.  Again, we have an overall minus sign difference to the sine-Gordon derived result simply because of the direction in which we integrated over the string, and because we integrated over only one spike we have only half of the sine-Gordon derived result.  Below we will see that, given an approximation, this same term also appears to capture the string's contribution to the angular momentum density at $\phit\neq 1$.

Computing $\Delta-J$ we get
\bea
(\Delta-J)|_{\phit=1}&=&T\int_{\sigma_{0}}^{\frac{\pi}{2}} \frac{s^{2}(1-r^{2})r'^{2}}{\rho^{2}}{\rm d}\sigma\nn\\
&=&T\kappa=\slambda
\eea
which is the finite result that we expect.  As noted in \cite{Hirano}, this contribution has arisen from the term $\Pi_{A}F_{\tau\sigma}$ appearing in the Hamiltonian, equation \eqref{H1}; the contribution is entirely from the electric flux on the world volume, that is from the giant magnon string attached to the giant graviton, for which the contribution to $\Delta-J$ is zero.  Computing the brane angular momentum $J_{{\rm B}}$ as
\bea
J_{{\rm B}}|_{\phit=1}&=&T\int_{\sigma_{0}}^{\frac{\pi}{2}}s^{2}\left[\frac{r^{2}\phit\rho^{4}}{\sqrt{D}}+\rho^{4}\right]\bigg|_{\phit=1}{\rm d}\sigma\nn\\
&=&T\int_{\sigma_{0}}^{\frac{\pi}{2}}s^{2}\rho^{2}{\rm d}\sigma,\qquad T=\frac{2}{\pi}N,\label{JBphit1}
\eea
and using the solution \eqref{phit1soln} gives
\be
J_{{\rm B}}\to N\mbox{ when }\sigma_{0}\to 0\mbox{ and }\kappa\to 0\mbox{ keeping  }\kappa{\rm cot}(\sigma)=1\mbox{ fixed},
\ee
the final condition being required to comply with the boundary condition $r(\sigma_{0})=1$, as found in \cite{Hirano}.  $J_{{\rm B}}=N$ is what we should expect from the BPS giant graviton. In \cite{McGreevy} McGreevy and Susskind treated the D3 brane with spherical symmetry and showed that the giants obey
\be
\frac{J_{{\rm B}}}{N}\leq\rho^{2}\label{Jbound}
\ee
where $\rho$ is the (constant) radius of the brane and BPS giants saturate the inequality.  If the brane is maximal then $\rho=1$.  Looking back at the integral \eqref{JBphit1} and being conscious of the behaviour of the $\phit=1$ solution which for $\kappa\to 0$ tends to a solution with a constant $\rho$ for $\sigma\in[\sigma_{0}\to 0,\frac{\pi}{2}]$ then we can see how we recover the BPS, spherical brane to which the string is attached, obtaining the same result as if we had taken $\rho^{2}$ out of the integral as a constant;
\bea
J_{{\rm B}}&=&\frac{2}{\pi}N\int_{\sigma_{0}}^{\frac{\pi}{2}}s^{2}\rho^{2}{\rm d}\sigma\nn\\
&\simeq&\frac{2}{\pi}N\rho^{2}\int_{0}^{\frac{\pi}{2}}s^{2}{\rm d}\sigma=\frac{2}{\pi}N\rho^{2}\cdot\frac{\pi}{2}=N\rho^{2}.
\eea
With $\rho=1$ we have $J_{{\rm B}}=N$.

With the energy of the brane given by an integral over the second term in equation \eqref{H2} for the total Hamiltonian density,
\bea
\Delta_{{\rm B}}|_{\phit=1}&=&T\int_{\sigma_{0}}^{\frac{\pi}{2}}\frac{s^{2}\rho^{4}}{\sqrt{D}}\bigg|_{\phit=1}{\rm d}\sigma\nn\\
&=&T\int_{\sigma_{0}}^{\frac{\pi}{2}}s^{2}\rho^{2}{\rm d}\sigma
\eea
which is equal to $J_{{\rm B}}|_{\phit=1}$ as given by \eqref{JBphit1}.  That is, the brane has
\be
\Delta_{{\rm B}}-J_{{\rm B}}=0\qquad\mbox{at }\phit=1.
\ee

Below we shall see that, given $\kappa\to 0$, the same is true for any $\phit\geq 0$.
\newline

Finally, we note that the field strength $\F$, given by \eqref{phit1soln}, as well as satisfying equation \eqref{AEOM}, satisfies Gauss's law on the world volume,
\be
{\rm div}_{S^{3}}F_{\tau a}=0,\qquad a\in\{\sigma,\sigma_{4},\sigma_{5}\}.
\ee
We have only the $\sigma$ component of the field strength and no dependence on the other world volume coordinates $\{\sigma_{4},\sigma_{5}\}$, hence
\be
{\rm div}_{S^{3}}=\frac{1}{\sqrt{h}}{\partial}_{a}\sqrt{h},\qquad h\equiv {\rm sin}^{4}(\sigma){\rm sin}^{2}(\sigma_{4}),\label{div}
\ee
where $h$ is the determinant of the metric on $S^{3}$, and the resulting factors of ${\rm sin}(\sigma_{4})$ cancel.  From \eqref{phit1soln}, the scalar $r$ should then satisfy
\be
\frac{1}{s^{2}}~\frac{\partial}{\partial\sigma}\left(s^{2}\frac{\partial}{\partial\sigma}~r\right)=0
\ee
which indeed it does.

This is analogous to the behaviour of the BIon spikes of Callan and Maldacena\cite{Callan:1997kz} where the linearised theory gave the same configurations as the fully non-linear approach; Maxwell theory was sufficient to construct the (BPS) objects of the Born-Infeld theory.  Similarly, the solutions presented in \cite{Sadri:2003mx} were constructed from a linearised approach to the giant graviton.  Being BPS the configuration is protected from receiving the corrections to the Born-Infeld approach that would normally be required when derivatives become large.

Defining the charge $Q$ as that seen by the Maxwell fields on the world volume, then while the net charge is of course zero, by performing a surface integral over the $S^{2}$ located at $\sigma$ we can find the charge of one of the point charges located at the poles to get
\bea
Q&=&\int_{S^{2}}\F{\rm d}A=s^{2}\F\Omega_{2}\nn\\
&=&4\pi\kappa.\label{Q}
\eea

\subsection{$\phit\neq 1$}

Away from $\phit=1$ it appears that the brane equations are no longer exactly solvable.  However upon making appropriate approximations a solution is available for any $\phit\geq 0$ that reproduces the Nambu string quantities derived above for both types of solution.  The first approximation we will take focusses our attention close to the poles of the deformed D-brane where the BIon spikes are to be found.  The second is to take $\kappa$ to be small, which as stated above means that we take the small string coupling, Nambu string limit.

First we examine the conserved quantities $J$ and $\Delta$. Using the $A_{\tau}$ equation \eqref{AEOM} to eliminate $\F$ we write the factor $\sqrt{D}$ as
\be
\sqrt{D}=\pm s^{2}\rho^{2}\Lambda\sqrt{\frac{r'^{2}+\rho^{4}}{s^{4}\rho^{4}+\kappa^{2}}},\qquad \Lambda\equiv\sqrt{\frac{1-\phit^{2}r^{2}}{1-r^{2}}}.\label{sqrtD}
\ee
If we take the string's contribution to the angular momentum to again be the ``$r'$ term'', and take the positive sign in front of $\sqrt{D}$, then we have for a single spike
\be
J=\int_{\sigma_{0}}^{\frac{\pi}{2}}\frac{Ts^{2}r^{2}\phit r'^{2}}{\sqrt{D}}{\rm d}\sigma=\int_{\sigma_{0}}^{\frac{\pi}{2}}\frac{Tr^{2}\phit r'^{2}}{\rho^{2}\Lambda}\sqrt{\frac{s^{4}\rho^{4}+\kappa^{2}}{r'^{2}+\rho^{4}}}{\rm d}\sigma.
\ee

Focussing on small $\sigma$ (or equivalently $\sigma$ close to $\pi$) we take $s^{4}\rho^{4}\ll\kappa^{2}$.  Around the spike we expect $r'$ to be large. In fact for $\phit=1$ we had $r'=\frac{\kappa}{s^{2}}$ so that when $s^{4}\rho^{4}\ll\kappa^{2}$ (and remember $0\leq\rho\leq 1$) then $|r'|\gg 1$.  If at $\phit\neq 1$ we expect there to remain a spiky solution that is continuously related to the $\phit=1$ case then we take this to remain true for $\phit\neq 1$.

An integral over $\sigma$ with lower limit of $\sigma_{0}$ will have to be cut off at some $\hat{\sigma}$ satisfying ${\rm sin}^{4}(\hat{\sigma})\rho^{4}\ll\kappa^{2}$.  For $\sigma>\sigma_{0}$ there should be vanishing contribution to quantities coming from the string / BIon spike.  Obviously a similar argument holds close to the other spike close to $\sigma=\pi$.

At leading order in this approximation then the factor $\sqrt{D}$ satisfies
\be
\frac{s^{4}\rho^{4}}{\kappa^{2}}\ll 1,\quad \frac{\rho^{4}}{r'^{2}}\ll 1,\quad\Ra\quad\sqrt{D}\approx s^{2}\rho^{2}\Lambda\frac{r'}{\kappa}
\ee
and the candidate string / BIon angular momentum becomes
\bea
J&=&\slambda\int_{\sigma_{0}}^{\frac{\pi}{2}}\frac{r^{2}\phit}{\rho^{2}\Lambda}\frac{r'^{2}}{\kappa}\sqrt{\frac{s^{4}\rho^{4}+\kappa^{2}}{r'^{2}+\rho^{4}}}{\rm d}\sigma\nn\\
&\approx&\slambda\int_{\sigma_{0}}^{\hat{\sigma}}\frac{r^{2}\phit}{\rho^{2}\Lambda}\frac{r'^{2}}{\kappa}\frac{\kappa}{r'}{\rm d}\sigma\label{Jcutoff}\\
&=&-\slambda\int_{\hat{r}}^{r_{{\rm U}}}\frac{r^{2}\phit}{\rho^{2}\Lambda}{\rm d}r
\eea
which is to be compared with equation \eqref{Js1} for the Nambu string's angular momentum.  The lower limit here is $\hat{r}\equiv r(\hat{\sigma})$.  Clearly if $\hat{r}\to 0$ then $J\to J_{{\rm s}}$, which will be the case if after taking $s^{4}\rho^{4}\ll\kappa^{2}$ we allow $\kappa\to 0$; we recover the Nambu string contribution in the zero string coupling (or large $N$ 't Hooft) limit which is by definition what we should expect.

So the leading order approximation to the angular momentum captures all of the detail of the Nambu string when $\kappa\to 0$, for any $\phit$, and as shown above this integral can be performed and matched with the results obtained explicitly from the string solutions.  Together with a similar simplification of the expression for the string / BIon energy $\Delta$, 
\be
\Delta\approx\Delta_{{\rm s}},\quad\mbox{to leading}\quad s^{4}\rho^{4}\ll\kappa^{2}\quad\mbox{and }\kappa\to 0\nn
\ee
we will obtain precisely the finite $J$ leading order corrections to $\Delta-J$ presented previously, such as in equation \eqref{EJs}, i.e.
\be
\Delta-J\approx\slambda\left\{1\pm\frac{4}{e^{2}}e^{-\frac{2\pi}{\sqrt{\lambda}}J}\right\}\label{EJb}
\ee
and indeed the full spectrum of $\Delta-J$ with $\phit$ discussed above.
\newline

Now we know the BIon spikes will return us the correct dispersion relations between the conserved quantities for any $\phit$, our next step is to find solutions that display the same behaviour as encountered in section 3.  The full equations appear intractable but we may be able to find approximate solutions given the above discussion.

To this end we once again eliminate $\F$ from the $r$ equation of motion \eqref{rEOM} using the $A_{\tau}$ equation \eqref{AEOM}.  Multiplying through by a factor of $\sqrt{D}$ and using the approximation $s^{4}\rho^{4}\ll\kappa^{2}$ we expand to first order to get a left hand side (LHS) of
\be
{\rm LHS}\approx\rho^{2}\Lambda s^{2} r'\left(1+\frac{1}{2}\frac{\rho^{4}}{r'^{2}}-\frac{1}{2}\frac{s^{4}\rho^{4}}{\kappa^{2}}\right){\rm d}_{\sigma}\left[\Lambda\left(1-\frac{1}{2}\frac{\rho^{4}}{r'^{2}}+\frac{s^{4}\rho^{4}}{\kappa^{2}}\right)\right].
\ee
Remembering that we expect $r's^{2}$ to be of order one, we by the same reasoning take $r''\lesssim r'^{2}$, then the remaining terms to first order in $\frac{s^{4}\rho^{4}}{\kappa^{2}}$ on the left hand side give
\bea
{\rm LHS}\approx\rho^{2}\Lambda r's^{2}\left\{\frac{rr'(1-\phit^{2})}{\Lambda\rho^{4}}\left(1-\frac{1}{2}\frac{\rho^{4}}{r'^{2}}+\frac{1}{2}\frac{s^{4}\rho^{4}}{\kappa^{2}}\right)-\left(-\frac{2rr'\rho^{2}}{r'^{2}}-\frac{\rho^{4}r''}{r'^{3}}\right)\right.\nn\\
\left.+\frac{\Lambda}{2}\left(\frac{4s^{3}{\rm cos}(\sigma)\rho^{4}}{\kappa^{2}}-\frac{4s^{4}rr'\rho^{2}}{\kappa^{2}}\right)+\left(\frac{1}{2}\frac{\rho^{4}}{r'^{2}}-\frac{1}{2}\frac{s^{4}\rho^{4}}{\kappa^{2}}\right)\frac{rr'\Omega^{2}}{\Lambda\rho^{4}}\right\}
\eea
where $\Omega^{2}\equiv 1-\phit^{2}$.  From the right hand side (RHS) we get
\bea
{\rm RHS}\approx-2r s^{2}\Lambda^{2}\frac{s^{4}\rho^{4}r'^{2}}{\kappa^{2}}\left(1+\frac{\rho^{4}}{r'^{2}}-\frac{s^{4}\rho^{4}}{\kappa^{2}}\right)\qquad\qquad\qquad\qquad\qquad\qquad\nn\\
+rs^{2}\left(\frac{r'^{2}\Omega^{2}}{\rho^{2}}-\rho^{4}\Lambda^{2}-\phit^{2}\rho^{4}\right)\pm 4rs^{2}\phit\rho^{4}s^{2}\Lambda\frac{r'}{\kappa}\left(1+\frac{1}{2}\frac{\rho^{4}}{r'^{2}}-\frac{1}{2}\frac{s^{4}\rho^{4}}{\kappa^{2}}\right).
\eea
The $\pm$ in front of the final term on the right hand side (which has been `squared out' elsewhere in the above expressions) comes from that in front of the $\sqrt{D}$, equation \eqref{sqrtD}, and is important.

Next we take our second approximation: $\kappa\to 0$.  Looking at the above, $\kappa$ appears only in negative powers and the leading term (in $\kappa$) is on the right hand side, being proportional to $\kappa^{-4}$.  On it's own we do not recover a realistic equation for $r'$ so we include the next to leading term, which is also on the right hand side and is proportional to $\kappa^{-3}$.  With the left hand side contributing only higher order terms in $\kappa$ we hence are led to another first order differential equation for $r$,
\be
0=\frac{2(s^{4}\rho^{4})^{2}\Lambda^{2}rr'^{2}}{\kappa^{4}}\mp\frac{4\rho^{4}s^{2}\phit\Lambda r'(\rho^{4}s^{4})}{2\kappa^{3}}
\ee
or
\be
r'=\pm\frac{\kappa}{s^{2}}\frac{\phit}{\Lambda}\label{rprime}
\ee
which generalises equation \eqref{phit1eqn} to $\phit\neq 1$.  At $\phit=1$ the factor $\Lambda=1$ and we recover the correct equation for $r'$.  The equation is again separable so that we get the \emph{implicit} solutions
\bea
1\leq \phit:&\qquad E\left(r\phit,\frac{1}{\phit}\right)-W^{2}F\left(r\phit,\frac{1}{\phit}\right)=c\pm\kappa{\rm cot}(\sigma)\label{soln1}\\
0\leq\phit\leq 1:&\qquad \frac{1}{\phit}E(r,\phit)=c\pm\kappa{\rm cot}(\sigma)\label{soln2}
\eea
where in \eqref{soln1} we define $W^{2}\equiv 1-\frac{1}{\phit^{2}}$.  As in the $\phit=1$ case a constant $c$ appears on the right hand side and a choice of sign.  The choice of sign is essential to be able to always maintain $r\geq 0$.  As in the $\phit=1$ case we still have the \emph{two} spikes, one of which occurs close to $\sigma=\pi$ where the ${\rm cot}(\sigma)$ is large and negative, hence requiring the choice of the minus sign.

\begin{figure}
\centering
\includegraphics[width=0.45\textwidth]{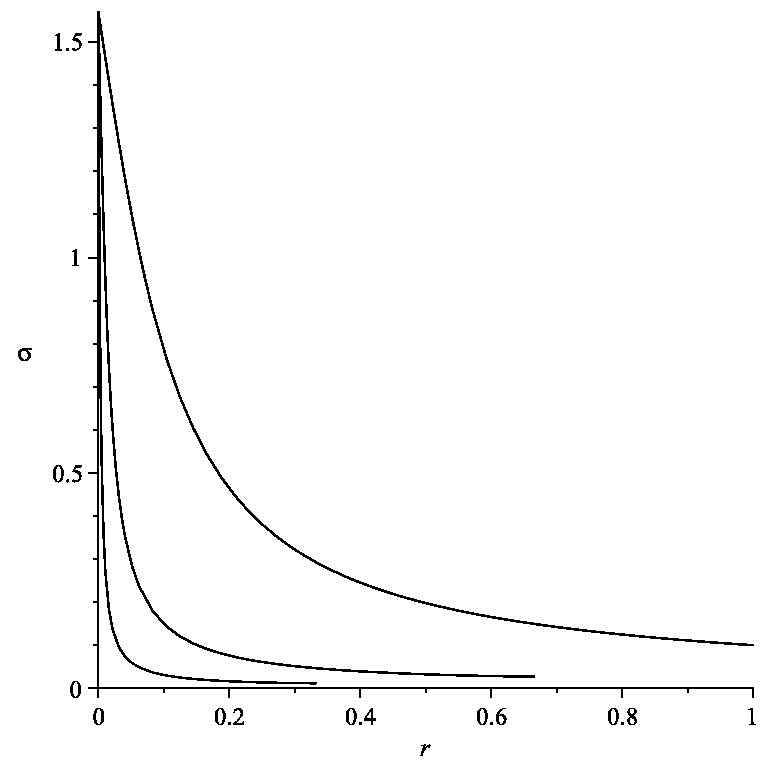}
\hspace{1.5in}\parbox{5in}{\caption{The $\phit\geq 1$ solution plotted at three different values of $\phit$ and three different values of $\kappa$.  We should in fact always have $\kappa$ small enough that the spike ($r>0$) is located at very small $\sigma$, but here we relax this condition so as to see the qualitative behaviour of the solution.  The values are $\phit=1$ and $\kappa=0.1$, $\phit=1.5$ and $\kappa=0.01$, $\phit=3$ and $\kappa=0.001$.\label{A}}}
\end{figure}

\begin{figure}
\centering
\includegraphics[width=0.45\textwidth]{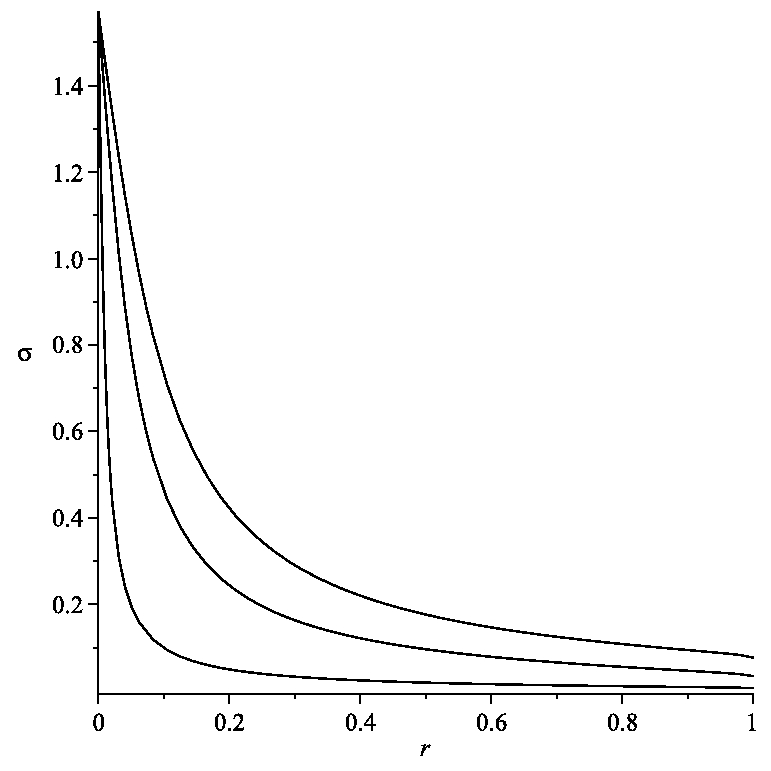}
\hspace{1.5in}\parbox{5in}{\caption{The $0\leq \phit\leq 1$ solution plotted at three different values of $\phit$ at $\kappa=0.1$.  Again $\kappa$ should be taken small enough that all of $r>0$ takes place at very small $\sigma$.  For any $\phit$ in the range the solution always obtains $r=1$, the equator of $S^{2}$, where the radius of the brane $R\phi$ shrinks to zero size.  Note that here $\kappa$ is held \emph{fixed}; as we decrease $\phi$ toward the stretched string at $\phit=0$ the solution becomes spike-like \emph{anyway}.\label{B}}}
\end{figure}

Again we encounter elliptic integrals, and the two solutions are the inverse modulus transform of one another\footnote{The Inverse modulus transform takes functions with elliptic modulus $k$ to functions with elliptic modulus $k^{-1}$. We use this when $k\geq 1$ so that the modulus used in the incomplete integrals is always less than or equal to 1.  For the incomplete elliptic integrals of the first and second kind respectively we we have
\be F\left(r\phit,\frac{1}{\phit}\right)=\phit F(r,\phit)\quad\mbox{and}\quad E\left(r\phit,\frac{1}{\phit}\right)=\frac{1}{\phit}\left[E(r,\phit)-(1-\phit^{2})F(r,\phit)\right].\nn\ee}.  While these are only implicit formulae for $r(\sigma)$ we can rearrange for $\sigma(r)$ to view their behaviour, even if inverted about $r=\sigma$.  Figures \ref{A} and \ref{B} plot $\sigma(r)$ for each solution.  Remembering our approximations we should keep $\kappa$ small, for which the solutions are very spiky, and we must only believe in ${\rm sin}^{2}(\sigma)\ll\kappa$, although as $\kappa\to 0$ the whole of the spike fits into this region, with $r'\to 0$ for all other values of $\sigma$.


For the $\phit\geq 1$ solution, with the lower limit of $r$, $\hat{r}\approx 0$, we obtain a string that has $0\leq r\leq \dfrac{1}{\phit}$ as expected; as we increase the angular velocity $\phit$ the end/cusp point retreats from the equator of the sphere at $r=1$.  Note that from equation \eqref{rprime} the spike has infinite gradient (in the world volume coordinate $\sigma$) at it's end/cusp point.   This is for any $\kappa$, in contrast to the $\phit=1$ solution for which this is true only at $\kappa\to 0$.  It should be remembered though that the solution took small $\kappa$ to begin with so that this change in behaviour between $\phit=1$ and $\phit>1$ is not visible in the limit.

For the $0\leq \phit\leq 1$ solution we obtain a string that has $0\leq r\leq 1$ as expected, always obtaining the point $r=1$ where the radius of the brane $R\rho$ shrinks to zero size.

Note that in both cases the points at or closest to the equator at $r=1$ are at the poles of $S^{3}$ on the world volume of the brane and would therefore seem to be the opposite end points of the stringy extensions of the brane.  This is in apparent contrast to the string solutions derived above where these points are in the middle of the string, the end points as determined by the ends of the ranges of the spacial world sheet coordinate being at the  `north' and `south' poles of the $S^{2}$.  However, the $\phit\geq 1$ string satisfies a free end point condition at it's cusp point so that it can equally well be thought of as two coincident strings with one Dirichlet condition each at $r=0$ and one Neumann condition each at the points closest to the equator.\footnote{This is similar to the case of the $J\to\infty$ giant magnon that must technically be thought of as part of a closed string but on its own passes for an isolated open string.}  In the $0\leq\phit\leq 1$ case the string could equally well be thought of as two separate strings satisfying both Dirichlet conditions at $r=0$ and Dirichlet conditions at $r=1$ in the vertical direction.  This would then appear to describe a string (pair) connecting a $Z=0$ giant graviton to a $Y=0$ giant graviton that meets the $S^{2}$ upon which the string moves only at the equator, $r=1$.
\newline

Given that our approximation allows us to consider almost all values of $r$ with $0<r\leq 1$ we may think about the constant $c$ once again.  Clearly, if $c=0$ we have the finite $J$ version of the boundary giant magnons attached to a maximal, BPS $Z=0$ giant graviton.  However, for $c\neq 0$ we would appear to be describing finite $J$ boundary giant magnons attached to $Z=0$ giant gravitons that are non-maximal and \emph{non}-BPS as the giant graviton itself must now orbit with $\phit\neq 1$.  From \cite{McGreevy} we get the bound \eqref{Jbound} which is saturated by BPS giants (for which $\Delta-J=0$) when $\phit=1$ only.
\newline

A simple form of the electric field $\F$ can be given at this level of approximation\footnote{i.e. the same level at which we return the Nambu string energy and angular momentum.}.  Expanding \eqref{F} to terms of order 1 in $s^{4}\kappa^{-2}$, which is just the leading order $\propto r'$, and using \eqref{rprime} we get 
\bea
\F&=&\Lambda r'+O\left(\frac{s^{2}}{\kappa}\right)\\
&\approx&\frac{\kappa}{s^{2}}~\phit
\eea
This is the $\phit=1$ result multiplied by $\phit$. Applying the divergence operator \eqref{div} we again satisfy Gauss's law,
\be
\frac{1}{s^{2}}\frac{{\rm d}}{{\rm d}\sigma}\left(s^{2}\cdot\frac{\kappa\phit}{s^{2}}\right)=0.
\ee
We can work out the magnitude of each of the point charges located at the poles as in \eqref{Q} once again to find
\be
Q=4\pi\kappa\phit.
\ee
We have an extra factor of $\phit$ compared with the $\phit=1$ case.  So changing $\phit$ changes the magnitude of the electric charges that are seen by the Maxwell fields on the world volume.
\newline

Turning to the angular momentum of the brane once again, the $\kappa\to 0$ limit squashes the spike into an infinitesimally small region close to $\sigma=0$ outside of which the brane tends to a constant radius of $R\rho=R$.  If we call $\sigma\in[\tilde{\sigma},\frac{\pi}{2}]$ the region in which the brane has a constant radius to a good degree of approximation, and will hence satisfy $\Delta_{{\rm B}}-J_{{\rm B}}\approx 0$, then we may also want to worry how the brane contribution $\Delta_{{\rm B}}-J_{{\rm B}}$ behaves in the region $\sigma\in[\sigma_{0},\hat{\sigma}]$ over which our solution is valid.  In particular we would like to see that it is also zero so that we maintain the picture of a Nambu string attached to a BPS giant graviton in the $\kappa\to 0$ limit.  To this end we argue as follows.

Substituting for $\F$ in $\sqrt{D}$ and examining the leading approximation,
\bea
\sqrt{D}&=&s^{2}\rho^{2}\Lambda\sqrt{\frac{r'^{2}+\rho^{4}}{s^{4}\rho^{4}+\kappa^{2}}}\nn\\
&\approx&s^{2}\rho^{2}\Lambda\cdot\frac{r'}{\kappa}\nn\\
&\approx&\phit\rho^{2}.
\eea
If we substitute this now into our expression for the brane angular momentum density in the region $\sigma\in[\sigma_{0},\hat{\sigma}]$ we find
\bea
J_{{\rm B}}&=&T\int_{\sigma_{0}}^{\hat{\sigma}}s^{2}\left[\frac{r^{2}\phit\rho^{4}}{\sqrt{D}}+\rho^{4}\right]{\rm d}\sigma\nn\\
&\approx& T\int_{\sigma_{0}}^{\hat{\sigma}}s^{2}\rho^{2}{\rm d}\sigma.
\eea
while the energy becomes
\be
\Delta_{{\rm B}}\approx \frac{T}{\phit}\int_{\sigma_{0}}^{\hat{\sigma}}s^{2}\rho^{2}{\rm d}\sigma
\ee
so that together with $0\leq\rho^{2}\leq 1$ we find
\bea
|\Delta_{{\rm B}}-J_{{\rm B}}|&\leq & T\bigg|\frac{1}{\phit}-1\bigg|\int_{\sigma_{0}}^{\hat{\sigma}}s^{2}{\rm d}\sigma\nn\\
&<&T\bigg|\frac{1}{\phit}-1\bigg|\frac{\hat{\sigma}^{3}}{3}.
\eea
Now, $\hat{\sigma}$ satisfied ${\rm sin}^{4}(\hat{\sigma})\rho^{4}\ll\kappa^{2}$ which will be true for any $\rho$ if ${\rm sin}^{4}(\hat{\sigma})\ll\kappa^{2}$, where $\kappa=\frac{\sqrt{\lambda}}{4N}$.  With ${\rm sin}(\hat{\sigma})\ll 1$ we have that $\hat{\sigma}^{3}\ll \kappa^{\frac{3}{2}}$ so that with $T=\frac{2}{\pi}N$ we have the bound
\be
|\Delta_{{\rm B}}-J_{{\rm B}}|\ll\frac{2}{3\pi}\bigg|\frac{1}{\phit}-1\bigg|\frac{\lambda^{\frac{3}{4}}}{4^{\frac{3}{2}}}\cdot\frac{1}{N^{\frac{1}{2}}}.
\ee
If we take $N\to \infty$, i.e. $\kappa\to 0$, with $\lambda$ fixed (and $\phit$ strictly non-zero) then $|\Delta_{{\rm B}}-J_{{\rm B}}|\to 0$ as would be expected of a BPS giant.

\section{Conclusion}
We began by finding the general solution to static sine-Gordon theory on the interval, producing two qualitatively different solutions depending upon the choice of a real-valued parameter.  We then mapped these to Nambu string solutions on $\mathbb{R}\times S^{2}$ where they correspond to a doubled up half of a folded, spinning string, or boundary giant magnon attached to $Z=0$ maximal giant gravitons.  There the choice of parameter turns out to be the choice of the string's angular velocity $\phit$ and the choice between $0\leq \phit\leq 1$ and $1\leq \phit$ separates two sets of qualitatively different open string solutions, one of which has been studied before\cite{Bak}.  The two types of solution converge at $\phit=1$ which from the sine-Gordon perspective is when the length of the interval $L$ on which the theory is defined diverges, recovering two infinitely separated copies of the boundary kink solutions and corresponding boundary giant magnons.

Taking $\phit>1$ we find the expected exponential corrections to the (finite) difference between string energy and angular momentum while for the new solutions at $\phit<1$ we find corrections of equal magnitude but opposite sign.  At $\phit=0$ we recover the static stretched string on $S^{2}$, supported against collapse by maintaining Dirichlet boundary conditions at each end.

We then moved to the world volume theory of the giant graviton itself in order to discover our boundary strings as BIon spike solutions of Born Infeld theory with a Chern-Simons potential.  Again, the known case of $\phit=1$ was recovered and with an appropriate pair of approximations that focussed on the rapidly varying part of the brane in the limit of small string coupling we found solutions that reproduced the behaviour of both sets of string solutions and at leading order were sure to possess the same angular momenta and energies.  It also followed from the solutions that in the zero string coupling limit, or large $N$ 't Hooft limit, the total brane configuration appeared to become the sum of a Nambu string and a BPS giant graviton.

We discussed the generalisation of the brane solutions that have non-zero constant $c$ at both $\phit=1$ and $\phit\neq 1$.  For small $c$ at least, where the brane description of the giants is valid, $\phit=1$ describes boundary giant magnons attached to a non-maximal, BPS, $Z=0$ giant graviton, while for $\phit\neq 1$ we have finite $J$ boundary giant magnons attached to a non-maximal, \emph{non}-BPS, $Z=0$ giant graviton.  It would be interesting to investigate non-maximal giant gravitons further and we intend to return to this topic in the future.
\newline\newline
\large{\bf{Acknowledgements}}\newline\newline
\small
A. Ciavarella and P. Bowcock would like to thank Douglas J. Smith for conversations relevant to this work.

\end{document}